\documentclass[11pt]{article}
\usepackage[sort&compress,square,comma,numbers]{natbib}

\usepackage[labelfont=bf]{caption} 
\usepackage[font={small}]{caption}   

\usepackage{subcaption}
\usepackage{amsmath}
\usepackage{amssymb}
\usepackage{url}
\usepackage{placeins}
\usepackage{textcomp}
\usepackage{graphicx}
\usepackage{float} 
\usepackage{chngpage}
\usepackage{authblk}

\usepackage[usenames,dvipsnames]{color}  

\usepackage{lineno}

\newcommand*\patchAmsMathEnvironmentForLineno[1]{%
  \expandafter\let\csname old#1\expandafter\endcsname\csname #1\endcsname
  \expandafter\let\csname oldend#1\expandafter\endcsname\csname end#1\endcsname
  \renewenvironment{#1}%
     {\linenomath\csname old#1\endcsname}%
     {\csname oldend#1\endcsname\endlinenomath}}% 
\newcommand*\patchBothAmsMathEnvironmentsForLineno[1]{%
  \patchAmsMathEnvironmentForLineno{#1}%
  \patchAmsMathEnvironmentForLineno{#1*}}%
\AtBeginDocument{%
\patchBothAmsMathEnvironmentsForLineno{equation}%
\patchBothAmsMathEnvironmentsForLineno{align}%
\patchBothAmsMathEnvironmentsForLineno{flalign}%
\patchBothAmsMathEnvironmentsForLineno{alignat}%
\patchBothAmsMathEnvironmentsForLineno{gather}%
\patchBothAmsMathEnvironmentsForLineno{multline}%
}

\usepackage[usenames,dvipsnames]{color}    % ADDED FOR COLOR NAMES
\usepackage{soul}   % SOUL Added for underlines
\setstcolor{red} % strikethrough color
\setul{}{1.5pt}  % line thickness

\newcommand{\beginsupplement}{%
	\setcounter{table}{0}
	\renewcommand{\thetable}{S\arabic{table}}%
	\setcounter{figure}{0}
	\renewcommand{\thefigure}{S\arabic{figure}}%
	\setcounter{page}{1}
	\renewcommand{\thepage}{S\arabic{page}}%
	\setcounter{section}{0}
	\renewcommand{\thesection}{S\arabic{section}}%
	\setcounter{equation}{0}
	\renewcommand{\theequation}{S\arabic{equation}}%
}

\usepackage{pifont}
\newcommand{\hollowstar}{\text{\ding{73}}}

\newcommand{\EMU}{$\text{EM}_\text{U}$ }
\newcommand{\EML}{$\text{EM}_\text{L}$ }

\begin{document}

\title{\textbf{Robust temporal pumping in a magneto-mechanical topological insulator}}

\author{Inbar Hotzen Grinberg$^1$, Mao Lin$^2$, Cameron Harris$^2$,
	\mbox{Wladimir A. Benalcazar$^2$}, Christopher W. Peterson$^3$,
	\mbox{Taylor L. Hughes$^{2\ast}$}, and Gaurav Bahl$^{1\ast}$\\
	\footnotesize{$^1$ Department of Mechanical Science and Engineering, $^2$ Department of Physics,}\\
	\footnotesize{$^3$ Department of Electrical and Computer Engineering,}\\
	\footnotesize{University of Illinois at Urbana-Champaign, Urbana, Illinois 61801, USA}\\
	\footnotesize{$^\ast$ To whom correspondence should be addressed; hughest@illinois.edu, bahl@illinois.edu} %\comment{$\Uparrow$ reformatted the title chunk for compactness}}
}

\date{}

	\vspace*{-2cm}
	{\let\newpage\relax\maketitle}
	
%\begin{linenumbers}
	\addtolength{\topmargin}{-0.75in}
	\addtolength{\textheight}{1in}
\begin{abstract}

The transport of energy through 1-dimensional (1D) waveguiding channels can be affected by sub-wavelength disorder, resulting in undesirable localization and backscattering phenomena. However, quantized disorder-resilient transport is observable in the edge currents of 2-dimensional (2D) topological band insulators with broken time-reversal symmetry. 
Topological pumps are able to reduce this higher-dimensional topological insulator phenomena to lower dimensionality by utilizing a pumping parameter (either space or time) as an artificial dimension.
Here we demonstrate the first temporal topological pump that produces on-demand, robust transport of mechanical energy using a 1D magneto-mechanical metamaterial. We experimentally demonstrate that the system is uniquely resilient to defects occurring in both space and time Our findings open a new path towards exploration of higher-dimensional topological physics with time as a synthetic dimension.

\end{abstract}

\maketitle

%%%%%%%%%%%%%%%%%%%%%%%%%%%%%%%%%%%%%%%%%%%%%%%%%%%%%%%%%%%%%%%%%%%%%%%%%%%%%%%%%%%%%%%%%%%%%%%%%%%%%%%%%%%%%%

The discovery that topological insulators host protected boundary states has spurred significant research on their metamaterial analogues due to attractive prospects in both science and engineering. 
A particularly important feature is the robust propagation that is observable in the chiral edge modes of 2D topological insulators having broken time-reversal symmetry, otherwise broadly known as Chern insulators \cite{Halperin1982,Buttiker1988,Haldane1988}. 
This class of systems includes integer quantum Hall insulators \cite{Klitzing1980}, the quantum anomalous Hall insulator \cite{Haldane1988,Chang2013}, and their metamaterial analogs \cite{Haldane2008,Raghu2008,Wang2009,Rechtsman2013,Susstrunk2015}, all of which can produce quantized transport even with significant disorder.

In this context, it was shown that periodic, adiabatic, spatio-temporal modulations of a 1D periodic potential can also produce quantized particle transport \cite{Thouless1983} where the number of particles pumped in one cycle is equal to the Chern number defined on the $(1 + 1)$-dimensional Brillouin zone spanned by momentum and time \cite{Thouless1982}. Thus, an adiabatic pumping process may be regarded as a dynamical manifestation of a Chern insulator in one higher dimension \cite{Haldane1988}, and as such is similarly topologically robust against disorder and defects \cite{Niu1984}.
Topological pumps have been implemented in a variety of systems including cold atomic gases \cite{Chien2015,Lohse2016,Nakajima2016,Lohse2018} and classical metamaterials \cite{Lu2014,Huber2016,Bertoldi2017}. Significant explorations in photonic metamaterials include using topological pumps to map the Berry curvature \cite{Xiao2010,Wimmer2017}, to demonstrate transport of a localized mode in a quasiperiodic waveguide array \cite{Kraus2012,Verbin2015}, and to probe a four dimensional quantum Hall effect \cite{Zilberberg2018}. 
However, to date, a temporally-controlled topological pump that produces on-demand, disorder-resilient transport has not been demonstrated in any metamaterial system.

Topological pumping can be understood as a consequence of the spectral flow property \cite{bernevigbook,Alexandradinata2011} of topological band structure. 
For the conventional 1D topological pump, the band structure evolves from topologically non-trivial to trivial and back during one pumping cycle, and crucially, reflection and time-reversal symmetries are \textit{not} preserved. For a system with open boundaries an integer number of eigenstates ``flow" from a lower energy band to an upper energy band, i.e. across the bulk energy gap, during this process, and the spatial profile of the flowing modes migrates from one end of the system to the other while carrying, e.g., charge, spin, or energy. 
Given a pumping protocol we can calculate a well-known topological invariant called the Chern number.
This invariant dictates the quantity of spectrally flowing modes, and hence the amount of, e.g., charge, spin, or energy that is \emph{robustly} transported across the (meta)material during one cycle of an ideal pump.  

From a tight-binding model perspective \cite{Qi2011}, a topological pump can be produced through spatio-temporal modulation of the on-site potentials and couplings between the constitutive elements of a metamaterial platform. However, not all cyclic spatio-temporal modulations generate non-vanishing Chern numbers.
Even if a protocol produces a Chern number, there are additional dynamical constraints for achieving robust transport. Namely, this process must be performed adiabatically to ensure that energy from the spectrally flowing states does not leak to the bulk bands.
At the same time, since physical systems have finite loss, the topological pump must complete a pump cycle faster than the decay time of the state being transported. 
For photonic implementations, the latter requirement necessitates extremely rapid modulation, which is technically very challenging. To date, a workaround has been to use space instead of time as the pumping parameter \cite{Kraus2012,Rechtsman2013, Verbin2015, Wimmer2017, Zilberberg2018,Lustig2019}. 
A time-controlled classical topological pump has remained elusive to date, and as a result, on-demand robust pumping of energy in a classical metamaterial has not yet been achieved.

In this work, we demonstrate a temporal topological pump using a 1D metamaterial composed of magnetically-coupled mechanical resonators. Pumping is achieved by replicating a 2D Chern insulator in one spatial dimension and one temporal dimension.
A non-contact approach is employed to produce the necessary modulations of the couplings and on-site potentials using permanent magnets and a high-permeability metal alloy mounted on a common rotating shaft. This system can, in principle, be ``hand cranked" to pump energy on-demand, in a manner reminiscent of an Archimedes Screw.
We experimentally demonstrate that mechanical energy can be robustly transported across the entire metamaterial in exactly one pumping cycle, as long as dynamical requirements listed above are met. 
We further demonstrate through a series of experiments that the topological pump is robust against disorder that may appear either in space or time.

\vspace{12pt}

%%%%%%%%%%%%%%%%%%%%%%%%%%%%%%%%%%%%%%%%%%%%%%%%%%%%%%%%%%%%%%%%%%%%%%%%%%%%%%%%%%%%%%%%%%%%%%%%%%%%%%%%%%%%%%%%

\begin{figure}[!th]
	\begin{adjustwidth}{-1in}{-1in}
		\vspace{0.5cm} 
		\makebox[\textwidth][c]{\includegraphics[width=1\textwidth]{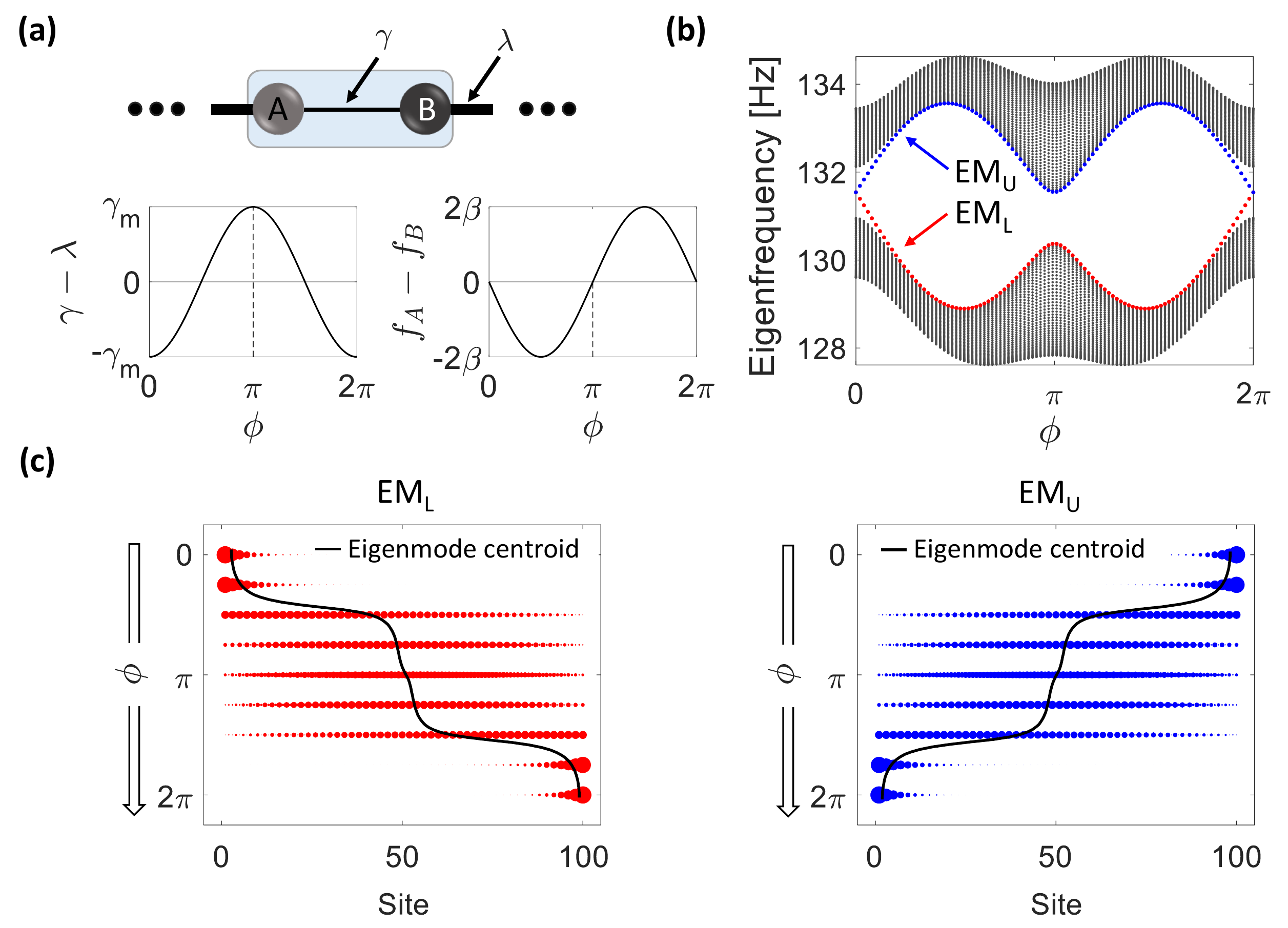}}
		\centering
		\caption{
			\textbf{Description of the topological pump.}
			\textbf{(a)} A unit cell of a dimerized 1D array (sub-lattice sites A, B) having on-site potential $f_A, f_B$, inter-cell coupling $\gamma$ and intra-cell coupling $\lambda$. The pump is produced by modulating the intra-cell coupling with a $\cos(\phi)$ trend, and modulating the on-site potentials with a $\sin(\phi)$ trend, where $\phi \in [0, 2\pi]$ is the pump phase.
			\textbf{(b)} Calculated band-structure for an array composed of 100 sites, as a function of the pump phase $\phi$. \EMU and \EML are the upper and lower trajectories over which the two edge modes evolve during the pump cycle. At $\phi = 0, 2\pi$ the system is in the topologically non-trivial phase while at $\phi = \pi$ the system is in the topologically trivial phase.
			\textbf{(c)} Evolution of the \EML (red) and \EMU (blue) eigenmodes during the pump cycle. Dots represent the magnitude of the eigenmode at each site. \EML is localized on the left edge at $\phi=0$ and transports to the right edge. In contrast, \EMU is transported from the right edge to the left edge. The solid black line represents the centroid of the eigenmode calculated through $\sum_{i}i\cdot\psi_{mi}$ where $\psi_{mi}$ is the normalized weight of the $m^{th}$ eigenmode at position $i$.
		}
		\label{fig:Concept}
	\end{adjustwidth}
\end{figure}

\newpage

We begin by developing the prescription of the topological pump on a 1D array of identical resonators having couplings with alternating strengths (dimerized) as depicted in Fig.~\ref{fig:Concept}a. 
Sub-lattices $A$ and $B$ correspond to the resonator positions inside a unit cell, with intra-cell coupling rate $\gamma$ and inter-cell coupling rate $\lambda$.
This system can be described through the well-known Su-Schreefer-Heeger model for polyacetylene \cite{Su1979,Su1980}, which informs us of the existence of two distinct phases in the presence of inversion symmetry.
The array is in a topologically non-trivial phase when $\gamma < \lambda$, protected by the approximate inversion or chiral symmetries, and is in a topologically trivial phase if $\gamma > \lambda$. A finite array composed of these unit cells in the non-trivial phase supports a mid-gap mode confined to each end of the chain.
For a translationally invariant chain with periodic boundary conditions the Bloch Hamiltonian of this system is written as
	\begin{equation}
		H (k_x)=(\gamma+\lambda\cos(k_x))\sigma_1 + \lambda\sin(k_x)\sigma_2,
		\label{eq:SSH_Hamiltonian}
	\end{equation}
where $k_x$ is momentum along the array, and $\sigma_1 = \bigl( \begin{smallmatrix}0 & 1\\ 1 & 0\end{smallmatrix}\bigr)$ and $\sigma_2 = \bigl( \begin{smallmatrix}0 & -i\\ i & 0\end{smallmatrix}\bigr)$ are the Pauli matrices.
The above system can now be modulated to produce the dynamic equivalent of a 2D Chern insulator \cite{Thouless1983} that is described by the momentum space Hamiltonian
	\begin{equation}
		H (k_x, \phi)=(\gamma+\lambda\cos(k_x)+\gamma_m\cos(\phi))\sigma_1 + \lambda\sin(k_x)\sigma_2 + \beta\sin(\phi)\sigma_3  ~.
		\label{ew:Chern_Hamiltonian}
	\end{equation}
Here we have introduced $\phi$ as an effective momentum in a second, synthetic dimension which in practice is the angular phase of the pumping cycle (which varies from $0$ to $2\pi$) and is proportional to time.
The modulation that introduces the Pauli matrix $\sigma_3 = \bigl( \begin{smallmatrix}1 & 0\\ 0 & -1\end{smallmatrix}\bigr)$ corresponds to odd-symmetric frequency modulation of the sublattices. 
This term breaks inversion symmetry during the pumping cycle and ensures that the Hamiltonian remains gapped throughout.
The parameters $\gamma_m$ and $\beta$ are the modulation depths of the coupling rates and the on-site potentials respectively.

Upon mapping from momentum space into real space, the Hamiltonian for the topologically pumped 1D array can be written as 
\begin{eqnarray}\begin{aligned}
	\label{eq:ChernHamiltonian}
		H (\phi) = \sum_n & \left( (\lambda-\gamma_m\cos\phi)a^\dagger_nb_n + \lambda b^\dagger_{n}a_{n+1}+h.c. \right. \\
				& \left. +\beta\sin\phi(a^\dagger_na_n-b^\dagger_nb_n) +\frac{\gamma_m}{2}\cos\phi(a^\dagger_na_n+b^\dagger_nb_n)\right)
	\end{aligned}
	\label{Hamiltonian}
\end{eqnarray}
where $a_n$($a^\dagger_n$) and $b_n$($b^\dagger_n$) are the annihilation and creation operators of the modes of interest on the two sub-lattice sites within the $n$-th unit cell. 
We achieve the above prescription by keeping the inter-cell coupling $\lambda$ fixed, while modulating the intra-cell coupling as $\gamma(\phi)=\lambda-\gamma_m\cos\phi$. Simultaneously, the on-site potentials are modulated as $\Delta f_A(\phi)=-\beta\sin\phi$ and $\Delta f_B(\phi)=+\beta\sin\phi$.
The prescribed modulations of the coupling and on-site potentials are graphically illustrated in Fig. \ref{fig:Concept}a. 
The last term in Eqn.~\ref{Hamiltonian}, which we did not include in Eqn. \ref{ew:Chern_Hamiltonian}, arises from behavior specific to our system \cite{Grinberg2019} as described in Supplement \S S1. However, since this term is identical on all sites, it does not change the eigenmodes or any robust properties of the topological pump, but only acts to shift the mode frequencies as a function of the phase in the pumping cycle. 

The pumping process can be illustrated as follows. Without loss of generality let the pumping phase $\phi=0$ or $2\pi$ represent the array in the topologically non-trivial phase, with two edge modes within the bandgap that are degenerate in frequency and positioned on opposite ends of the chain (Supplement \S S4.5). %\ref{sec:suppl_Band_structure}).
We identify these modes as the lower edge mode (EM$_\text{L}$) and the upper edge mode (EM$_\text{U}$), due to the paths in frequency they follow during the pumping cycle. 
As $\phi$ evolves away from $0$, both \EML and \EMU become dispersive and merge into the bulk with \EML decreasing in frequency and \EMU increasing in frequency (Fig. \ref{fig:Concept}b). 
At exactly mid cycle $\phi=\pi$ the array recovers inversion symmetry but is now in the topologically trivial phase. 
As $\phi$ continues evolving towards $2\pi$, the edge modes re-emerge from the bulk bands, and have now migrated to the opposite physical ends of the array from where they started (Fig.~\ref{fig:Concept}c). 
Since $H(\phi)$ is gapped for all $\phi$, we can calculate the Chern number of the pumping cycle, which for our system is 1 (see Supplement \S S2). This means that the pumping process is topologically protected and \EML and \EMU are topologically robust to disorder and smooth changes of system parameters.
%

%%%%%%%%%%%%%%%%%%%%%%%%%%%%%%%%%%%%%%%%%%%%%%%%%%%%%%%%%%%%%%%%%%%%%%%%%%%%%%%%%%%%%%%%%%%%%%%%%%%%%%%%%%%%%%%%

\begin{figure}[p]
	\begin{adjustwidth}{-1in}{-1in}

		\vspace{-3cm} 

		\makebox[\textwidth][c]{\includegraphics[width=1.3\textwidth]{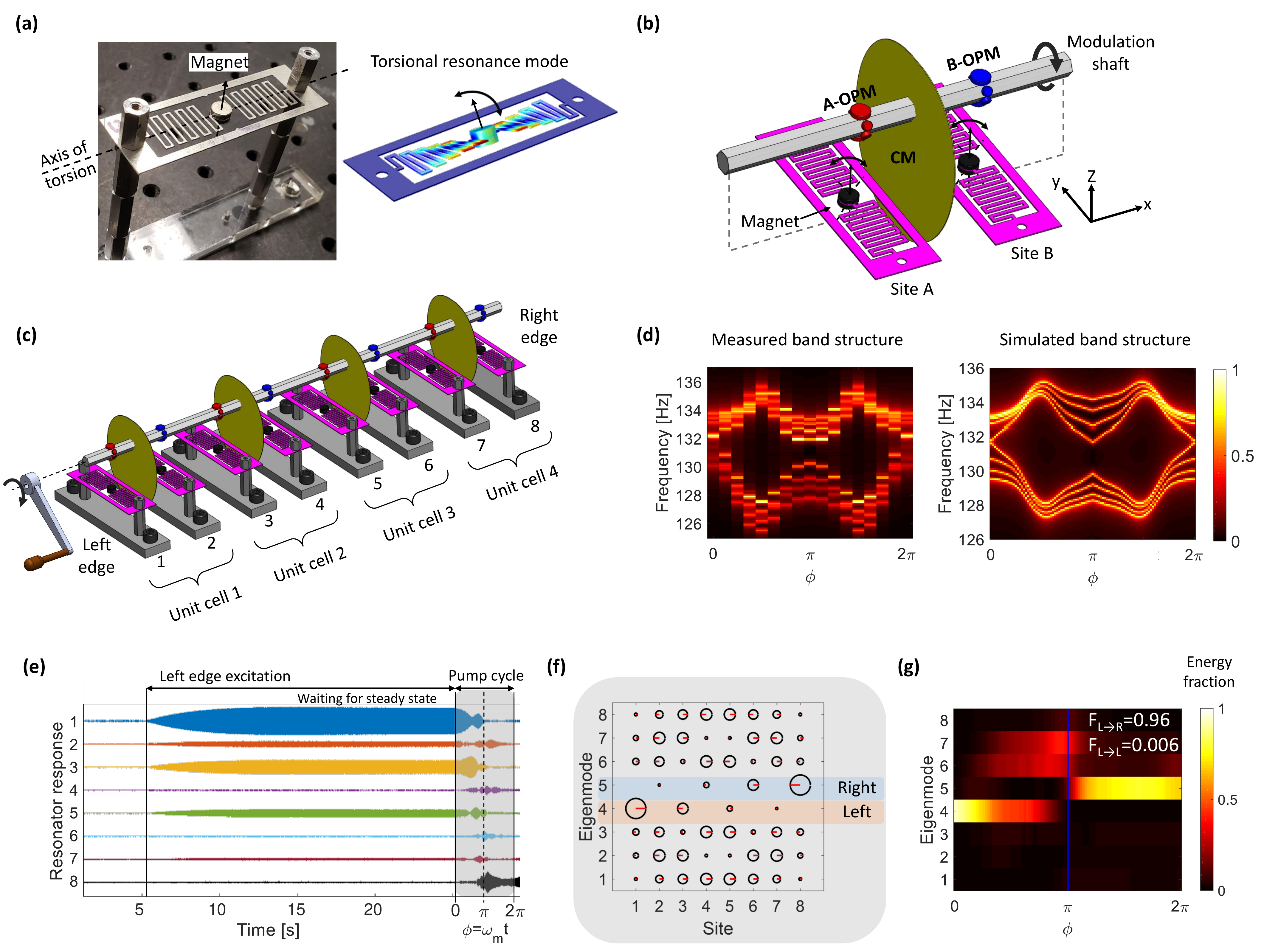}}
		\centering
		\caption{
			\textbf{Magneto-mechanical system description.}
			\textbf{(a)} Photograph of an individual magnet-loaded mechanical resonator and its simulated torsional resonance mode.
			\textbf{(b)} One unit-cell is composed of two mechanical resonators (Site A and Site B) that are coupled through their mutual magnetic interaction. Angular rotation $\phi$ of the modulation shaft produces the topological pump, by simultaneously inducing coupling modulation (CM) and on-site potential modulations (OPM). The CM is generated with an off-axis high-permeability metal alloy sheet (Supplement \S S4.3) while the OPM utilizes permanent magnets on the shaft (Supplement \S S4.4). See also Supplementary Animation 1.
			\textbf{(c)} All experiments use an array of 4 unit-cells. Mechanical coupling between the resonators is only achieved through the magnetic interaction.
			Rotation of the shaft from $\phi=0$ to $2\pi$ is motorized, though in principle a crank handle (illustrated) could be used to produce the pump.
			\textbf{(d)} Band structure is obtained quasi-statically by averaging the normalized mechanical impedance spectrum measured at each resonator, for discrete shaft angles $\phi$. Experimental measurements (left) and simulations (right) confirm the existence of edge states at $\phi = 0, 2\pi$ and an open bulk band gap for all $\phi$. The simulation includes non-ideal effects such as next nearest neighbor coupling (details in Supplement \S S4.5).
			\textbf{(e)} In a typical pumping experiment, we perform simultaneous time-domain measurements of the deflection of each resonator in the array. The array is driven only at site 1, as a result of which primarily the left edge mode (mode 4 in (f)) is excited. After stabilization, the drive is turned off and the pump (shaft rotation) is immediately activated. The example shows a pumping process with rotation rate $\omega_m \approx 0.3$ Hz.
			\textbf{(f)} Calculated eigenmodes of the eight resonator array at $\phi=0$. Circle size corresponds to the magnitude of the eigenmode while the red line indicates the corresponding phase (0 to the right, $\pi$ to the left). 
			The localized topological modes on the left and right edges (modes 4, 5) are highlighted. 
			\textbf{(g)} Temporal heat map representing the modal energy fractions ($E_\text{mode \#}$) during the pumping process, obtained by projecting the vibrational motion in (e) on to the eigenmode basis in (f). We see the transport of mechanical energy from the left edge mode (mode 4) to the right edge mode (mode 5) in the course of one pump cycle. Fidelity parameter $F_{L \rightarrow R} \approx 1$ (defined in text) indicates successful pumping.
		}
		\label{fig:System}
	\end{adjustwidth}
\end{figure}

We implemented the topological pump using an array of magnetically-coupled mechanical resonators.
Each resonator (Fig.~\ref{fig:System}a) is identically fabricated from waterjet-cut aluminum. A  neodymium magnet is bonded onto the central platform and serves both as the resonant mass as well as the mechanism by which adjacent resonators are magnetically coupled. The serpentine spring provides the restoring torque and sets the frequency for the torsional resonance mode at 132.4 Hz. 
The magnetically induced torque between the resonator dipoles couples their rotational degrees of freedom, and is used to produce the topological band structure. The coupling rate decays cubically with distance and can also be modified by placing high-permeability material between the resonators.
Details on the magnetic interaction and the equations of motion specific to this system are presented in the Supplement \S  S1. 
The typical -3 dB bandwidth of our resonators is $\Delta f \approx 0.38$ Hz which implies a decay time constant of $\tau = 1/(\pi \Delta f) \approx 0.85$ sec. This timescale is not sufficient for an experimental observation of topological pumping since, as we discuss later (and in the Supplement \S S3) the adiabaticity timescale of the system is around 1.6 sec.
Therefore, for each resonator we implement an anti-damping circuit that provides a velocity-dependent feedback force to increase the decay time to 3.5 sec (details in Supplement \S S4.2).

A single unit cell of the array is comprised of two resonators as shown in Fig.~\ref{fig:System}b, corresponding to sub-lattice sites A and B.
The experiment employed four unit cells as illustrated in Fig.~\ref{fig:System}c. A photograph of the experimental setup is provided in the Supplement \S S4.1. 
We physically implemented the pump cycle using a rotating shaft whose angular rotation directly represents the pump phase $\phi$ and which can be in essence `cranked' whenever the pump needs to be activated.
A clockwise (cw) rotation of the shaft corresponds to increasing $\phi$ from $0$ to $2\pi$, while counter-clockwise (ccw) rotation corresponds to decreasing $\phi$ from $2\pi$ to $0$.
This shaft is designed so that its rotation simultaneously produces the required coupling modulations and the required frequency (on-site potential) modulations without any physical contact with the resonator array, by means of only magnets and ferromagnetic materials.
The on-site frequency modulations are implemented by leveraging the magneto-static spring effect \cite{Grinberg2019}. This effect originates from the angular displacement-dependent torque acting on the magnetic harmonic oscillator in a non-uniform background magnetic field. Here we place permanent magnets on appropriate facets of the shaft to induce the required $\phi$-dependent frequency modulation (see Supplement \S S4.3 for details). 
We similarly modulate the intra-cell resonator coupling $\gamma$ using high-permeability mumetal sheets mounted off-axis on the modulation shaft (Fig.~\ref{fig:System}b). During rotation these sheets enter the gap between the site A and B resonators and change the coupling as a function of $\phi$. The specific geometry of the coupling modulation sheets is discussed in the Supplement \S S4.4.
%
%An animation of how the coupling-modulators and on-site potential modulators act during rotation of the shaft is provided in Supplementary Video S1.
%
Each resonator is equipped with a Hall sensor that measures its angular displacement. All eight resonators in the array are measured simultaneously so that both the magnitude of displacement and the relative phase can be known. 
During experiments, the excitation of the mechanical motion of any resonator is achieved using a sinusoidal magnetic field produced by a drive coil placed nearby.

\vspace{12pt}

We begin the experiment by performing a quasi-static characterization of the band structure of the magneto-mechanical states through the pumping cycle. 
The magneto-mechanical susceptibility (density of states) for any site in the array can be measured by actuating with a coil and measuring the calibrated angular displacement as a function of excitation frequency. These susceptibility measurements are then averaged over all resonators to produce a visualization of the mechanical density of states, as a function of shaft angular position, i.e., pump phase $\phi$.
The experimentally measured band structure for the system composed of 4 unit-cells is shown in Fig.~\ref{fig:System}d, and matches very well with the theoretical band structure, which we modeled using couplings to nearest and next-nearest neighbors.
This quasi-static measurement confirms that the band gap remains open throughout the pump cycle, and that mid-gap topological edge modes are present at $\phi=0$ and $2\pi$. 
We provide additional discussion on this band structure in the Supplement \S S4.5.

We can now demonstrate the dynamic pumping cycle and show that the energy in the left edge mode is robustly transported across the array to the right edge.
We start each pumping experiment by exciting the left edge resonator at the frequency of the topological edge mode. The excitation continues until a steady state response is reached. The excitation is then turned off and 
the modulation shaft is immediately activated to undergo one complete rotation, thereby evolving $\phi$ from $0$ to $2\pi$. 
An example of a typical measured angular displacement as a function of time for all 8 resonators is presented in Fig.~\ref{fig:System}e. In the representative example shown, mechanical energy is observed to transport across the array and localize on the right edge (resonator \#8).

Of key interest to this study is to quantify the localization of the mechanical energy through the pumping cycle, with special attention placed on the two edge modes in the topologically non-trivial configuration at $\phi=0$ and $\phi=2\pi$. 
We therefore establish the $\phi=0$ eigenmode set (Fig.~\ref{fig:System}f) as a convenient basis in which we can analyze the modal energy distribution. 
Here we can also define an energy fraction for a mode ($E_\text{mode \#}$) as the fraction of total mechanical energy in the array projected onto the selected mode.
The energy fraction for all 8 modes is traced throughout the pump cycle using overlapping 0.25 sec time segments (see Supplement \S 4.6) -- an example temporal heat map from an experimental measurement is presented in  Fig.~\ref{fig:System}g. 
At the beginning of the pumping cycle the mechanical energy primarily sits on basis mode 4, corresponding to the left edge mode, while at the end of the cycle the energy transports to basis mode 5 which corresponds to the right edge mode.
To further quantitatively analyze the pumping cycle we define a transport fidelity parameter $F_{L \rightarrow R} = E_5( \phi=2\pi)/E_4(\phi=0)$ as the ratio between energy fraction in the right edge mode at the end of the cycle, and the energy fraction in the left edge mode at the beginning of the cycle. 
This parameter quantifies how much of the initial energy in the left edge mode has transported across the array, and is a measure of the performance of the pump.
Similarly, the parameter $F_{L \rightarrow L} = E_4(\phi=2\pi)/E_4(\phi=0)$ indicates how much mechanical energy remains in the left edge mode at the end of the cycle. 
In an ideal pump cycle we expect $F_{L \rightarrow R} = 1$ and $F_{L \rightarrow L} = 0$. 
For the specific example shown in Fig.~\ref{fig:System}g the measured transport fidelity is $F_{L \rightarrow R} = 0.96$ demonstrating a successful pumping cycle. As we discuss below, the transport fidelity remains very high even in the presence of disorder as long as the adiabatic timescale is respected.

\begin{figure}[!hp]
	\begin{adjustwidth}{-1in}{-1in}
		\makebox[\textwidth][c]{\includegraphics[width=1.3\textwidth]{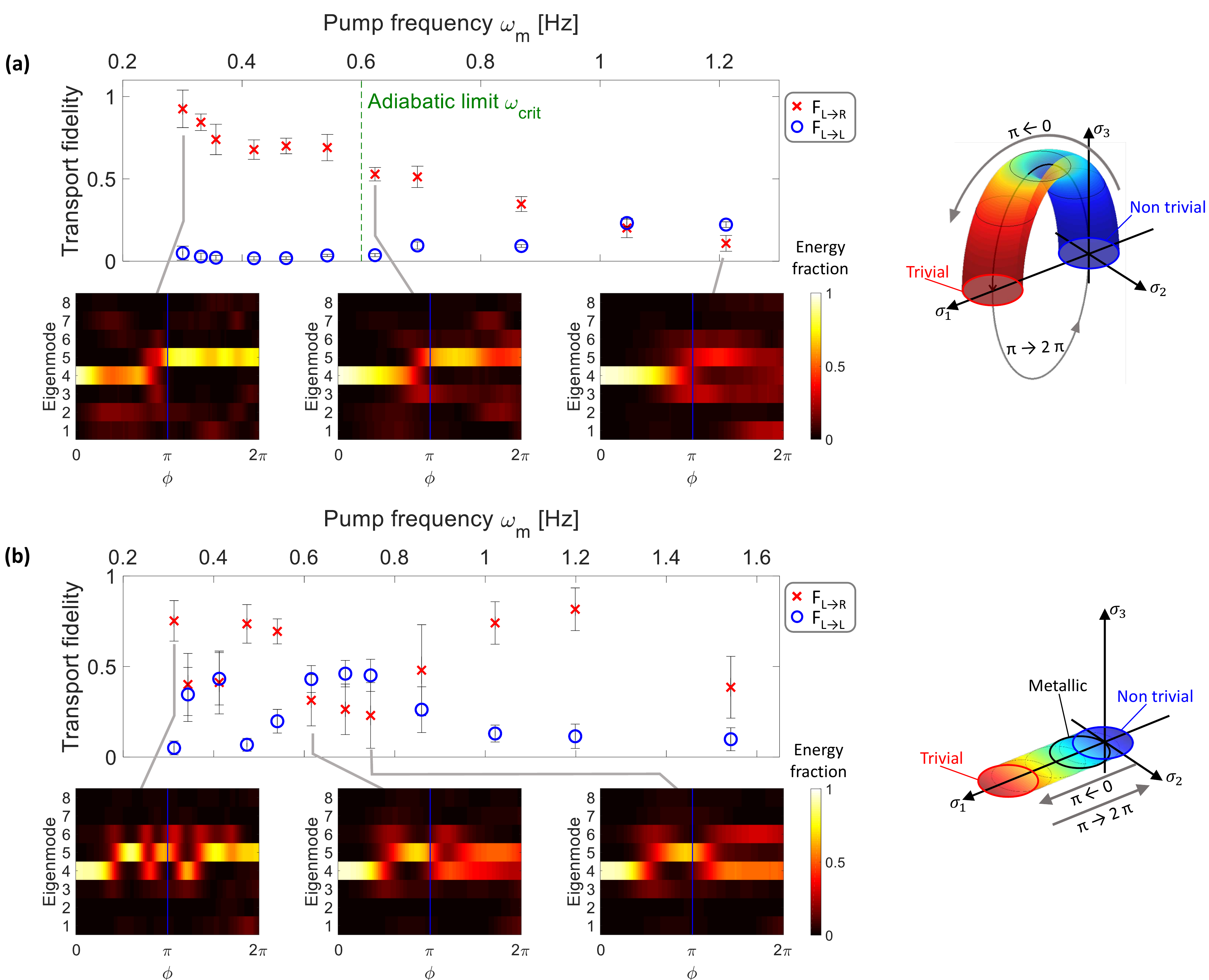}}
		\centering
		\caption{
			\textbf{Experimental measurement of transport fidelity and exploration of the adiabatic regime.}
			\textbf{(a)}
			We experimentally tested a range of pump frequencies $\omega_m$ traversing the critical threshold $\omega_\text{crit}$. For $\omega_m < \omega_\text{crit}$ the transport fidelity $F_{L \rightarrow R}$ remains high, and diminishes for $\omega_m > \omega_\text{crit}$. Each data point corresponds to the average of fidelity measurements from 10 consecutive pumping experiments and error bars show the standard deviation between experiments. 
			The insets show representative temporal heat maps of the energy fraction of each basis mode in Fig.~\ref{fig:System}f. In the adiabatic regime, mechanical energy robustly transports from mode 4 (left edge) to mode 5 (right edge), while in the non-adiabatic regime the energy disperses amongst the bulk modes. 
			The Pauli matrix basis representation of the pumping Hamiltonian is presented on the right, and shows how the Hamiltonian evolves during the pumping cycle between non-trivial and trivial phases while avoiding closing the band gap.
			\textbf{(b)} We also tested the non-adiabatic case where band gap closes twice during the cycle, as shown by the Pauli matrix basis representation on the right. In this case the Chern number cannot be used to characterize the process. Mechanical energy is instead seen to oscillate between the two edge modes, and therefore pumping is strongly dependent on timing.
			Supporting simulations for both cases are provided in the Supplement \S S4.7.
		}
		\label{fig:results}
	\end{adjustwidth}
\end{figure}

\vspace{12pt}

Having demonstrated on-demand temporal pumping in the magneto-mechanical resonator array, we turn to illustrate the importance of adiabaticity.  The pumping process timescale is characterized by the frequency $\omega_m$ which also corresponds to the angular rotation rate $d\phi/dt$ of the shaft.
Intuitively, the adiabatic condition is such that the frequency of  the Hamiltonian modulations during the pumping process must be smaller than the frequency gap between between a given eigenmode (EM$_\text{L}$ or EM$_\text{U}$ in our case) and the rest of eigenmodes, to mitigate transitions between the modes. Quantitatively, we calculate a critical pump frequency $\omega_\text{crit} \approx 0.6$ Hz above which the adiabaticity of the system breaks down \cite{Privitera2018} (calculation in Supplement \S 3). We expect that it is only in the adiabatic regime that the pumping process is characterized by non-vanishing Chern number of 1 (Supplement \S 2) and is therefore topologically protected. 
To show the breakdown of adiabaticity we experimentally measured values of $F_{L \rightarrow R}$ and $F_{L \rightarrow L}$ as a function of increasing pump frequency $\omega_m$. 
Fig.~\ref{fig:results}a presents the measured fidelities averaged over 10 consecutive experiments. We observe that pumping is achieved ($F_{L \rightarrow R}$ approaches 1) below the theoretically calculated $\omega_\text{crit} \approx 0.6$ Hz, and diminishes past this threshold.
The example insets show how the energy transports from mode 4 (left edge) to mode 5 (right edge) in the adiabatic pumping regime, but disperses amongst other bulk modes in the non-adiabatic pumping regime.

A limiting case where the adiabatic condition necessarily breaks is if the band gap closes at some point during the pumping cycle.
In this situation, there is not a well-defined Chern number associated with the pumping process and the reliable transfer of energy between edge states requires precise timing since it is a result of the coupling between the two edge modes instead of a topological pump.
To illustrate this non-adiabatic process, we modify the modulation shaft to turn off the resonator frequency modulations and only retain the coupling modulations. As a result, the Hamiltonian for the system (Eqn. 2) no longer contains the $\sigma_3$ term, and the band gap closes twice during the pump cycle, i.e., the system transits through a (bulk) conducting phase, as illustrated in Fig.~\ref{fig:results}b (see also Supplement Fig.~S9).
Once again, Fig.~\ref{fig:results}b presents experimental measurements of the transport fidelity as a function of pump frequency $\omega_m$. The values of $F_{L \rightarrow R}$ and $F_{R \rightarrow L}$ are seen to be irregular with no clear regime of pump frequency separating high and low values. Moreover, the example insets show that the mechanical energy oscillates between the two edge modes (modes 4 and 5) during the cycle confirming that the transport of mechanical energy is timing-dependent.

The results presented in Fig.~\ref{fig:results} are for cw rotation of the modulation shaft ($\phi$ increasing) i.e. pumping along \EML. An additional set of experiments with ccw rotation ($\phi$ decreasing) implying a pumping trajectory along \EMU are presented in the Supplement \S S5.1 along with supporting simulations in Supplement \S S4.7. As expected, ccw pumping also confirms the same adiabaticity characteristics.

\begin{figure}[!ht]
	\begin{adjustwidth}{-1in}{-1in}
		\makebox[\textwidth][c]{\includegraphics[width=0.91\textwidth]{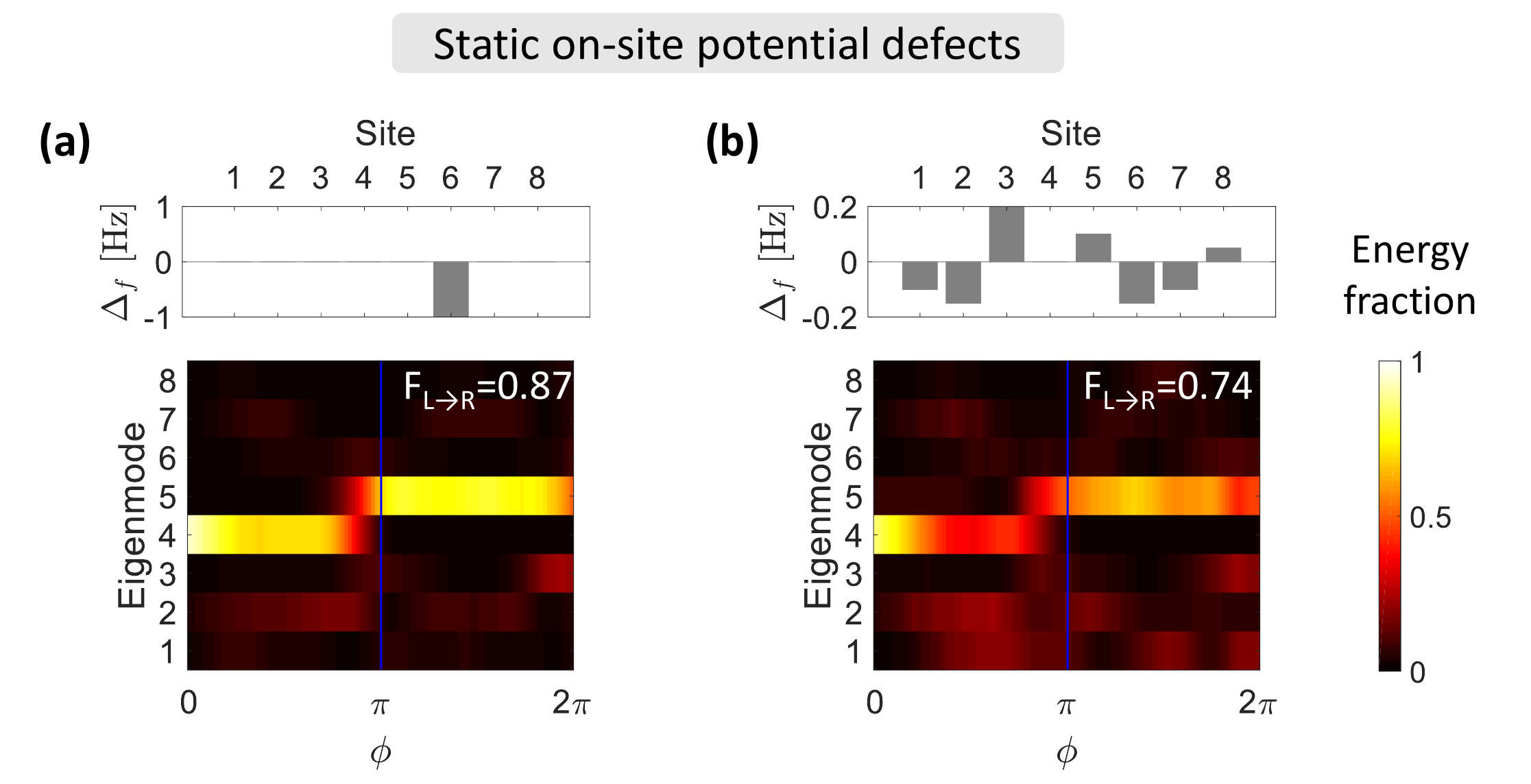}}
		\centering
		\caption{
				\textbf{Pumping in the presence of on-site potential disorder.} 
				\textbf{(a)} A single resonator is detuned by 1 Hz. 
				\textbf{(b)} Resonators are randomly detuned by $\pm 0.2$ Hz, corresponding to $\sim 10$\% of the band gap. 
				In both cases the temporal heat map of energy fractions shows robust pumping with high transport fidelity.
		}
		\label{fig:Defect}
	\end{adjustwidth}
\end{figure}

\begin{figure}[!hp]
	\begin{adjustwidth}{-1in}{-1in}

		\vspace{-3cm} 
		
		\makebox[\textwidth][c]{\includegraphics[width=1.1\textwidth]{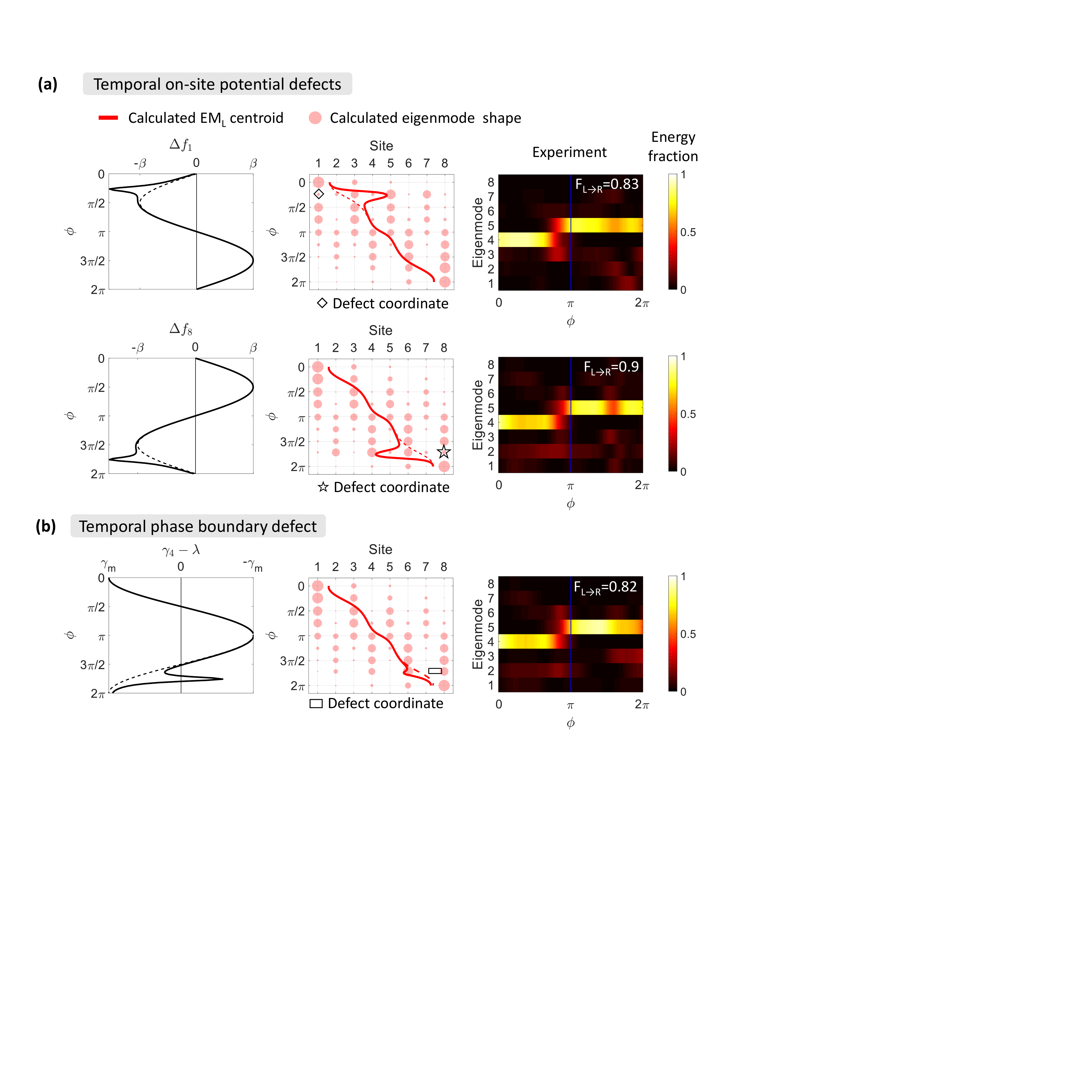}}
		\centering
		\caption{
			\textbf{Implementation of spatio-temporal defects.}
			Since the governing Hamiltonian of the topological pump matches that of a Chern insulator, pumping is robust to defects along its pseudo-edge in space and time. We use the centroid of the mode (middle column) to track this pseudo-edge and place defects along its path.
			\textbf{(a)} 
			On-site potential defects at the space-time coordinates marked by the $\diamond$ and \hollowstar.
			The frequencies $f_1$ and $f_8$ are modified away from the prescribed trajectory during the pumping cycle. We calculate (middle panel) how the mode centroid trajectory is modified for these defects. For both cases the experimentally measured modal energy fraction (right panel) exhibits high fidelity values indicating that the pump remains robust. 
			\textbf{(b)} 
			We also explored the intrusion of a trivial phase at the boundary of the equivalent 1+1D Chern insulator. This is implemented by modifying the intra-cell coupling rate $\gamma_4$ at the 4th unit cell, between sites 7 and 8, with space-time coordinates marked by $\sqsubset\!\sqsupset$ in the middle panel. Here $\gamma_4$ deviates from the prescribed trajectory during the pumping process (inter-cell coupling rate $\lambda$ remains fixed). Once again, the experimentally measured modal energy fraction shows high fidelity pumping in spite of the trivial phase intrusion.
		}
		\label{fig:DefectTemporal}
	\end{adjustwidth}
\end{figure}

\vspace{12pt}

Since the adiabatic pump is characterized by a non-vanishing Chern number, we expect the process to be robust to defects that deform the band structure but do not close the band gap.
One class of static defects that satisfy this criterion is the detuning of on-site potential, for which we present two specific examples in Fig.~\ref{fig:Defect}. The first example has a single resonator frequency detuned by 1 Hz. The second example uses a randomized detuning of $\pm 0.2$ Hz, corresponding to $\sim$10\% of the system band gap.
Results from pumping experiments show high transport fidelity $F_{L \rightarrow R}$ for both cases. A wide range of additional examples are presented in the Supplement \S S5.2 and exhibit consistent robustness against non-time-varying on-site potential disorder.

As mentioned previously, the Hamiltonian describing this system (Eqn.~\ref{eq:ChernHamiltonian}) is effectively that of a Chern insulator with one real spatial dimension and one synthetic frequency dimension.
Therefore, the system should exhibit robustness against defects that deform the pseudo-space-frequency edge of the equivalent 1+1D Chern insulator.
To find this pseudo-edge, we analyzed the spatio-temporal trajectory of the \EMU and \EML modes by visualizing their centroids in space (resonator site) and time (pump phase $\phi$) as shown in Fig.~\ref{fig:Concept}c for 100 sites and Supplement Fig.~S13 for 8 sites.
This visualization reveals the approximate space-time coordinates of mechanical energy through the pump cycle and helps to position the defects. At the beginning and the end of the cycle the mechanical energy is mostly localized near the left and right edge respectively, while in the middle of the cycle the mechanical energy propagates through the bulk.
Based on this analysis, we experimentally implemented on-site potential defects to coincide with the \EML centroid trajectory at site 1 at time $\phi = \pi/4$, and at site 8 at time $\phi = 7\pi/4$.
The defects were designed to be a simple momentary perturbation of on-site potential (i.e., $\Delta f_1, \Delta f_8$ respectively in Fig.~\ref{fig:DefectTemporal}a) by modifying the permanent magnets on the shaft at the corresponding sites and phase angle $\phi$. 
The experimental measurements presented in Fig.~\ref{fig:DefectTemporal}a show that pumping fidelity $F_{L \rightarrow R}$ remains very high in both of the above cases.
Heuristically the effective chiral edge mode simply avoids the defects and robustly pumps across the array without backscattering.
A series of additional experiments implementing this type of temporal on-site potential defect are shown in the Supplement \S S5.2, Fig.~S15 and Fig.~S16.
Finally, we also implemented a coupling defect that mimics a phase boundary deformation, i.e. an intrusion of a trivial phase into the bulk of the 1+1D Chern insulator. This type of defect would act to deform the edge of the effective 2D system, and we would still expect the chiral edge state to adapt and travel around the new boundary geometry. This defect was implemented by momentarily increasing the intra-cell coupling $\gamma_4$ at the 4th unit cell at time $\phi = 7\pi/4$. 
An intuitive visualization of this defect is presented in the Supplement Fig.~S17.
Experimental results from pumping in this array (Fig.~\ref{fig:DefectTemporal}b) show that mechanical energy temporarily re-localizes in the penultimate unit-cell, but the overall transport fidelity $F_{L \rightarrow R}$ at the end of the cycle remains very high.
All these experiments confirm the unique form of robustness of this topological pump against defects occurring in both space and time.

\vspace{12pt}

Linear waveguides are a foundational technology that enable modern systems for communications, sensing, and fundamental science. However, disorder that is frozen-in during fabrication, or appears dynamically in the form of fluctuations, can result in undesirable scattering \cite{Marcuse:1969, John:1988, Kim:2017} and localization \cite{Schwartz:2007} in these systems.
While spatial topological pumps can address these concerns, the introduction of time as a pumping parameter offers unprecedented control and reconfigurability over the transport of energy in space \cite{Chien2015,Lohse2016} and even in frequency \cite{martin2017topological}.
Moreover, the use of alternative pumping protocols or multiple incommensurate temporal drives can potentially open up a wide configuration space \cite{Peng2018,Kolodrubetz2018}, allowing the synthesis of larger Chern numbers for increased pumping capacity\cite{schroter2018topological, nielsen2018higher, song2016route, skirlo2015experimental, skirlo2014multimode}, the generation of higher Chern numbers in higher synthetic dimensions\cite{Petrides2018}, and the exploration of dynamic phase transitions between topological phases in time \cite{Zurek2005,Vajna2015,Solnyshkov2016}.

%\end{linenumbers}
\section*{Acknowledgments}
We acknowledge funding support from the National Science Foundation Emerging Frontiers in Research and Innovation NewLAW program (grant EFMA-1627184), an Office of Naval Research Director of Research Early Career Grant (grant
N00014-16-1-2830), and a National Science Foundation Graduate Research Fellowship for CWP. This work was supported in part by the Zuckerman STEM Leadership Program.

\newpage

\bibliographystyle{naturemag}

\pagebreak

\beginsupplement

\begin{center}
	\textbf{\large Robust temporal pumping in topological magneto-insulator: Supplementary Material}
\end{center}

\section{Equations of motion and derivation of system Hamiltonian}
\label{sec:suppl_eq_motion_svaa}

In this section we will derive the equations of motion of the magneto-mechanical resonator array, and reproduce the Hamiltonian presented in Eqn. 3 of the main text.

To model the magnetic interaction between the resonators, we consider each magnet-loaded resonator as a point dipole. This is heuristically acceptable as long as the distance between magnets is greater than their largest geometrical dimension.  
Generally, any magnetic dipole $\vec{m}$ placed within any magnetic field $\vec{B}$ feels a torque given by $\vec{\tau}=\vec{m}\times\vec{B}$. 
Each point dipole is a source to a non-uniform magnetic field, and as a result the torque acting on dipole $A$ due to dipole $B$ is given by \cite{Landecker1999}
\begin{equation}
\tau_\text{AB}=\frac{\mu_0}{4\pi r^5}(3\vec{m}_B\times(\vec{m}_A\cdot\vec{r})\vec{r}-r^2(\vec{m}_B\times\vec{m}_A)).
\label{eq:torque}
\end{equation}
Here $\mu_0$ is the magnetic permeability of free space, $\vec{m}_A,\vec{m}_B$ are the two magnetic dipoles, and $\vec{r}$ is their relative position (pointing from $A$ to $B$).

\begin{figure}[h]
	\begin{adjustwidth}{-0.5in}{-0.5in}
		\makebox[\textwidth][c]{\includegraphics[width=0.6\textwidth]{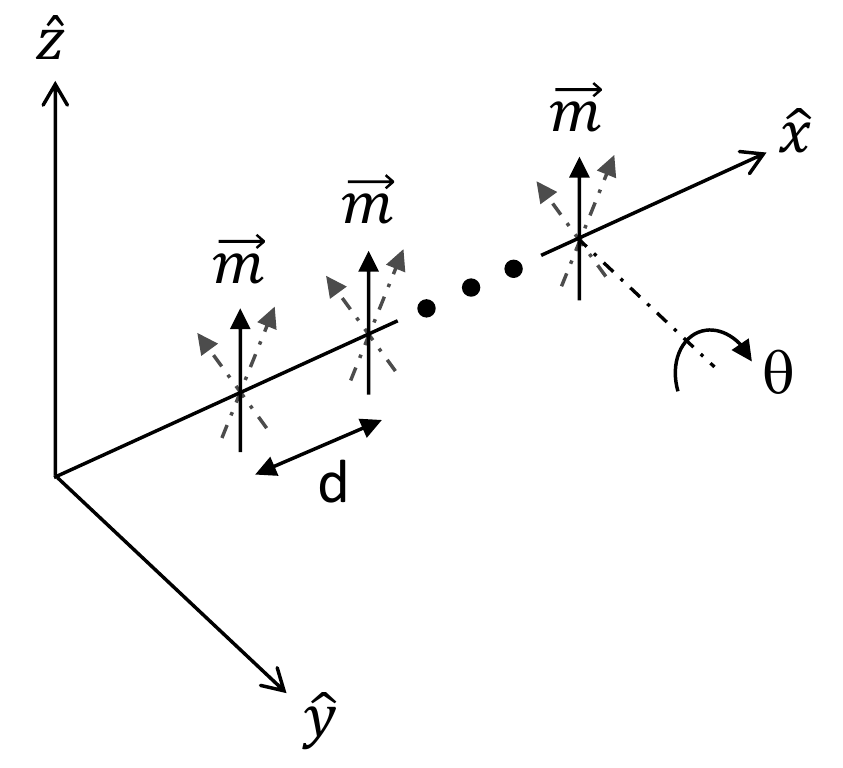}}
		\centering
		\caption{
			Our system is composed of a linear array of magnetic resonators, each having a single rotational degree of freedom. This diagram accompanies the discussion in the text.}
		\label{fig:Supp_dipole_diagram}
	\end{adjustwidth}
\end{figure}

In our experimental setup all magnetic dipoles are oriented along the $\hat{z}$ axis at rest, i.e. $\vec{m}=m\hat{z}$,  and are spaced only along the $\hat{x}$ axis, such that $\vec{r}=d\hat{x}$. 
Each dipole has a single rotational degree of freedom $\theta$ around the $\hat{y}$ axis, and we assume all dipoles are identical so that $m_A=m_B=m$ as illustrated in Fig. \ref{fig:Supp_dipole_diagram}. We substitute these conditions into Eqn. \ref{eq:torque} and use a small angle approximation \cite{Grinberg2019} to simplify the torque acting on dipole $A$ due to dipole $B$ to the following
\begin{equation}
\tau_\text{AB}=\frac{2\mu_0 m^2}{4\pi d^3}\left(\frac{\theta_A}{2}+\theta_B\right).
\end{equation}
Since the torque $\tau_{AB}$ on A is linearly dependent on the displacement $\theta_A$, it corresponds to a linear spring term that acts in parallel with the mechanical spring $\kappa$ of that resonator. In addition, we note a second coupling term that applies a torque on $A$ due to the angular displacement of $B$. The resulting coupled equations of motion for the $A$ and $B$ site resonators can then be written as:
\begin{equation}
\begin{cases}
\begin{array}{l}
\ddot{\theta_A}+c\dot{\theta_A}+(\omega_0^2-\frac{\gamma}{2})\theta_A-\gamma\theta_B=0\\
\ddot{\theta_B}+c\dot{\theta_B}+(\omega_0^2-\frac{\gamma}{2})\theta_B-\gamma\theta_A=0
\end{array}
\end{cases}
\label{eq: Second_Order}
\end{equation} 
where
\begin{equation}
c=\frac{b}{I}\ ;\ \omega_0^2=\frac{\kappa}{I}\ ;\ \gamma=\frac{2\mu_0 m^2}{4\pi d^3 I}.
\end{equation}
Here $b$ is the viscous damping coefficient, $I$ is the rotational moment of inertia, and $\omega_0$ is the natural resonance frequency of the mechanical resonator when isolated in space. The parameter $\gamma$ is the magnetic interaction which induces both a magnetic spring effect and coupling between adjacent resonators \cite{Grinberg2019}. 
The spring effect induced by this magnetic interaction results in the last term in Eqn. 3 in the main text. This effect is identical to all resonators in the array and therefore only induces a uniform frequency shift and will not result in closing of a band gap.

Now, given that $A,B$ are the sublattice sites of a dimerized array with $n$ unit cells having a fixed inter-cell coupling $\lambda$, a modulated intra-cell coupling $\gamma(\phi)=\lambda-\gamma_m cos\phi$, and modulated resonance frequencies $\Delta f _\text{A,B}(\phi)=\pm \beta sin(\phi)$ we can write the equations of motion of the modulated array
\begin{equation}
\medmuskip=0mu
\begin{cases}
\begin{array}{l}
\label{eq:suppl_EOM_magneto}			
\ddot{\theta}_{n,A} + c\dot{\theta}_{n,A} + (\omega_{r}^2(\phi)-\beta \sin \phi)\theta_{n,A} - \lambda \theta_{n-1,B} - \gamma(\phi)\theta_{n,B} = 0 , \\
\ddot{\theta}_{n,B} + c\dot{\theta}_{n,B} + (\omega_{r}^2(\phi)+\beta \sin \phi)\theta_{n,B} - \lambda \theta_{n+1,A} - \gamma(\phi) \theta_{n,A} = 0 
\end{array}
\end{cases}	
\end{equation}	
where the effective resonance frequency of each resonator in the array is
\begin{equation}
\omega_r^2(\phi)=\omega_0^2 -\lambda + \frac{\gamma_m}{2} \cos \phi .
\end{equation}

Next, we invoke the slowly varying envelop approximation (SVEA) to reduce the order of the equations and write the system Hamiltonian.
SVEA is the assumption that the envelope of the time domain amplitude changes slowly compared to the period of oscillations. Typical resonance frequencies of our resonators are around 130 Hz, while the modulation rate (which indicates the envelope) is $\sim$ 1 Hz. This means there are two orders of magnitude difference between the timescale of the resonator oscillations and the time varying envelope, which justifies the use of SVEA.
We now denote the oscillations of each resonator in the $n^{th}$ unit cell as $\theta_{n,\eta}$ with $\eta=A,B$, and assume that they take the following harmonic form
\begin{eqnarray}\begin{aligned}
\label{eq:SVAA1}
\theta_{n,\eta}(t) = x_{n,\eta}(t) e^{i\omega t} +c.c.
\end{aligned}\end{eqnarray}
where $\omega$ is the frequency of the external drive, and $x_{n,\eta}(t)$ is the amplitude of oscillation. Taking the time derivatives of Eqn. \ref{eq:SVAA1} we obtain
\begin{eqnarray}\begin{aligned}
\label{eq:SVAA2}
\dot{\theta}_{n,\eta}(t) &= \dot{x}_{n,\eta}(t) e^{i\omega t} + i\omega x_{n,\eta}(t) e^{i\omega t} +c.c. \\
\ddot{\theta}_{n,\eta}(t) &= \ddot{x}_{n,\eta}(t) e^{i\omega t} + 2i\omega \dot{x}_{n,\eta}(t) e^{i\omega t} -\omega^2x_{n,\eta}(t) e^{i\omega t} + c.c.  
\end{aligned}\end{eqnarray}
Under SVEA we set $\ddot{x}_{n,\eta}(t)=0$, and substitute Eqn. \ref{eq:SVAA1}-\ref{eq:SVAA2} into Eqn.~\ref{eq:suppl_EOM_magneto}. The equations of motion become
\begin{equation}
\medmuskip=0mu
\begin{cases}
\begin{array}{l}
\label{eq:EOM_magneto_slow_wave}
2i\omega\dot{x}_{n,A}-\omega^2x_{n,A} + c(\dot{x}_{n,A}+i\omega x_{n,A}) + \left(\omega_{r}^2-\beta\sin\phi\right)x_{n,A} - \lambda x_{n-1,B} - \gamma(\phi) x_{n,B} =0 \\
2i\omega\dot{x}_{n,B}-\omega^2x_{n,B} + c(\dot{x}_{n,B}+i\omega x_{n,B}) + \left(\omega_{r}^2+\beta\sin\phi \right)x_{n,B} - \lambda x_{n+1,A} - \gamma(\phi)  x_{n,A} =0
\end{array}
\end{cases}	
\end{equation}	
Rearranging the equations yields
\begin{eqnarray}\begin{aligned}
(i+\frac{c}{2\omega})\dot{x}_{n,A} &= \left(\frac{\omega^2-\omega_{r}^2+\beta\sin\phi}{2\omega}-\frac{ic}{2}\right) x_{n,A} + \frac{\lambda}{2\omega}x_{n-1,B} + \frac{\gamma(\phi)}{2\omega}x_{n,B} \\
(i+\frac{c}{2\omega})\dot{x}_{n,B} &=  \left(\frac{\omega^2-\omega_{r}^2-\beta\sin\phi }{2\omega}-\frac{ic}{2}\right)x_{n,B} + \frac{\lambda}{2\omega}x_{n+1,A}  + \frac{\gamma(\phi)}{2\omega}x_{n,A}
\end{aligned}\end{eqnarray}
%$\omega_r^2=\omega_0^2-\frac{1}{2}\gamma_\text{1,0}-\frac{1}{2}\lambda$
%
Since excitation frequency $\omega \approx \omega_0$, and losses are small $c \approx 0$, we obtain the dynamical equations 
\begin{eqnarray}\begin{aligned}
i\dot{x}_{n,A} &=\frac{1}{2\omega}\left( \left(\frac{\gamma_m \cos(\phi)}{2}+\beta\sin\phi \right) x_{n,A} + \lambda x_{n-1,B} + (\lambda-\gamma_m\cos(\phi)x_{n,B}\right) \\
i\dot{x}_{n,B} &=\frac{1}{2\omega}\left( \left(\frac{\gamma_m \cos(\phi)}{2}-\beta\sin\phi \right) x_{n,B} + \lambda x_{n+1,A} + (\lambda-\gamma_m\cos(\phi)x_{n,A}\right)
\end{aligned}\end{eqnarray}
This set of first order equations reveals all the couplings in the system and can be mapped to a Hamiltonian given by
\begin{eqnarray}\begin{aligned}
H = \frac{1}{2\omega} \sum_n&\left( (\lambda-\gamma_m\cos\phi)a^\dagger_nb_n + \lambda b^\dagger_{n}a_{n+1}+h.c. \right. \\& \left. +\beta\sin\phi(a^\dagger_na_n-b^\dagger_nb_n) +\frac{\gamma_m}{2}\cos\phi(a^\dagger_na_n+b^\dagger_nb_n)\right)
\end{aligned}
\label{eq:HamiltonianReal}
\end{eqnarray}
where $a_n$($a^\dagger_n$), $b_n$($b^\dagger_n$) are the standard annihilation and creation operators. Thus we recover Eqn. 3 of the main text.

%%%%%%%%%%%%%%%%%%%%%%%%%%%%%%%%%%%%%%%%%%%%%%%%%%%%%%%%%%%%%%%%%%%%%%%%%%%%%%%%%%%%%%%%%%%%%%%%%%%%%%%%%%%%%%%%%%%%%
\newpage

\section{Calculation of Chern number}
\label{sec:suppl_Cal_Chern}

\begin{figure}[H]
	\begin{adjustwidth}{-0.5in}{-0.5in}
		\makebox[\textwidth][c]{\includegraphics[width=0.5\textwidth]{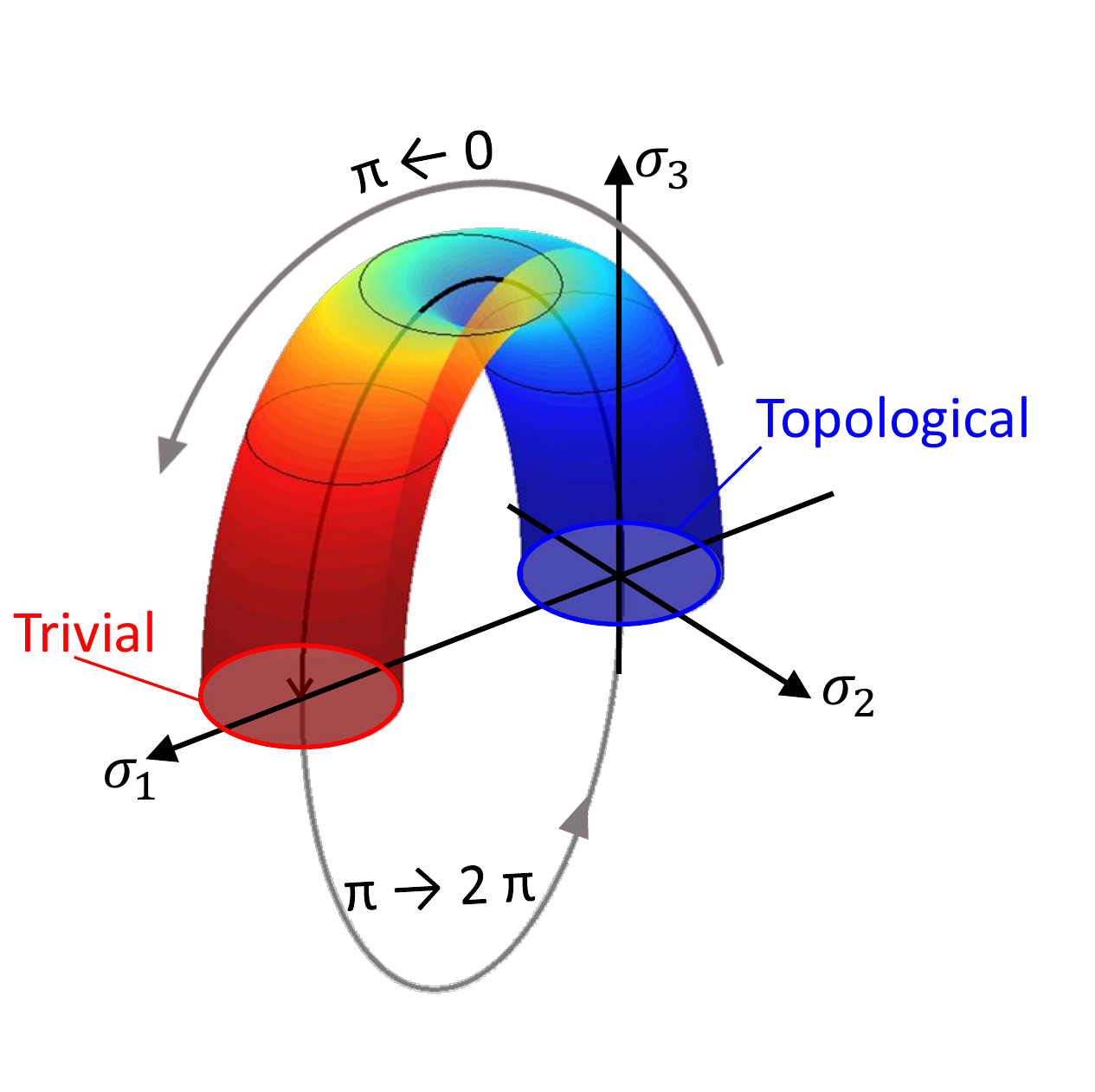}}
		\centering
		\caption{
			A visual representation of the vector $\hat{d}$ which describes the modulated Hamiltonian of the topological pump (Eqn. \ref{eq:dvec}) in the Pauli matrix space.}
		\label{fig:Supp_kSpaceFM}
	\end{adjustwidth}
\end{figure}

For a one dimensional dimerized array subjected to periodic modulations, the most general Hamiltonian takes the form
\begin{eqnarray}\begin{aligned}
H(k,\phi) = \epsilon(k,\phi) + \sum_{i=1}^3d_i(k,\phi)\sigma_i,
\end{aligned}\end{eqnarray}
where the quasi-momentum and the angular position $(k,\phi)$ effectively define a two-dimensional parameter space. 
Here, $\sigma_i$ represents the Pauli matrices $\sigma_1 = \bigl( \begin{smallmatrix}0 & 1\\ 1 & 0\end{smallmatrix}\bigr)$, $\sigma_2 = \bigl( \begin{smallmatrix}0 & -i\\ i & 0\end{smallmatrix}\bigr)$, $\sigma_3 = \bigl( \begin{smallmatrix}1 & 0\\ 0 & -1\end{smallmatrix}\bigr)$, and $d_i(k,\phi)$ are the components of a vector $\hat{d}$ that describes the Hamiltonian in the Pauli matrix space.
As we are following a state adiabatically to complete the pumping process, $\epsilon(k,\phi)$ only contributes to the dynamical phase \cite{griffiths_schroeter_2018}, and does not affect the Chern number. This is also demonstrated by the Chern number definition 
\begin{eqnarray}\begin{aligned}
\label{eq:Chern_num_def}
\nu = \frac{1}{4\pi}\int dk\,d\phi\,\hat{d}\cdot\frac{\partial\hat{d}}{\partial k}\times\frac{\partial\hat{d}}{\partial \phi},
\end{aligned}\end{eqnarray}
which is independent of $\epsilon(k,\phi)$. 
Writing the Hamiltonian for our system in momentum space, we find that the vector $\hat{d}$ is given by
\begin{eqnarray}\begin{aligned}
\hat{d} = \left\{\lambda+\lambda\cos(k)-\gamma_m \cos(\phi) , \lambda\sin(k) , \beta \sin(\phi)\right\}.
\label{eq:dvec}
\end{aligned}\end{eqnarray}
In the Pauli matrix space this vector is represented by a torus enclosing the origin as illustrated in Fig. \ref{fig:Supp_kSpaceFM}. Due to the  $\sigma_3$ component of the vector, the path of the Hamiltonian never crosses through the origin which is the singularity where the band gap closes.

Following the definition given in Egn. \ref{eq:Chern_num_def} the Chern number is analytically calculated for our system as
\begin{eqnarray}
\nu =
\begin{cases}
-1, & \lambda<-\frac{\gamma_m}{2} \\
\hspace{0.25cm} 0, &  -\frac{\gamma_m}{2}<\lambda<\frac{\gamma_m}{2} \\
\hspace{0.25cm} 1, & \lambda>\frac{\gamma_m}{2}
\end{cases}
\end{eqnarray}
In our experimental system the values of the parameters are in the range of $\lambda>\frac{\gamma_m}{2}$ and therefore the Chern number is 1.
\vspace{1mm}

For the non-adiabatic process demonstrated in the main text the $\sigma_3$ term is zero since there is no on-site potential modulation, and the vector $\hat{d}$ becomes
\begin{eqnarray}\begin{aligned}
\vec{d} = \left\{\lambda+\lambda\cos(k)-\gamma_m \cos(\phi) , \lambda\sin(k) , 0 \right\}.
\label{eq:dVecNon}
\end{aligned}\end{eqnarray}
\begin{figure}[H]
	\begin{adjustwidth}{-0.5in}{-0.5in}
		\makebox[\textwidth][c]{\includegraphics[width=0.5\textwidth]{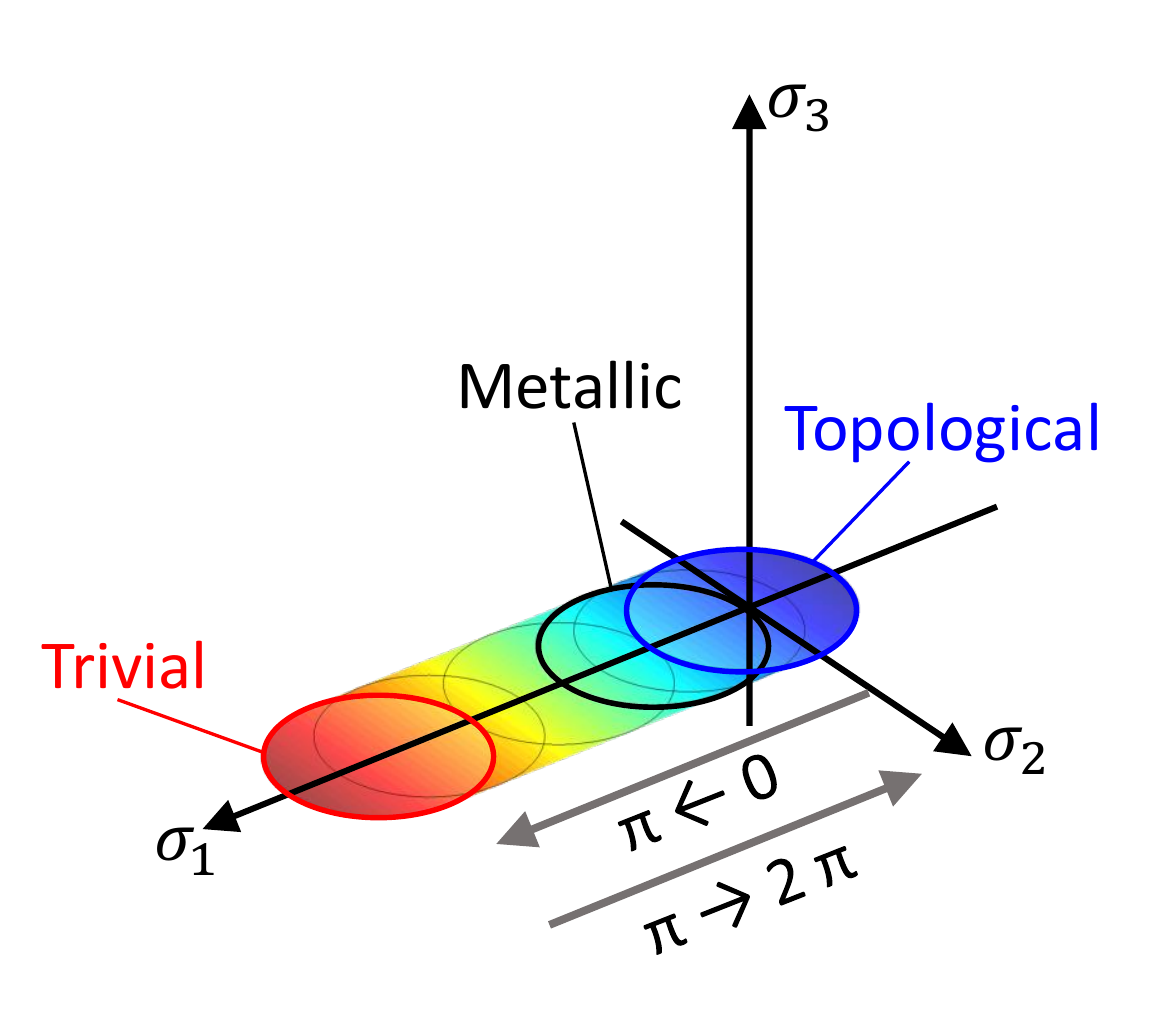}}
		\centering
		\caption{
			A visual representation of the vector $\hat{d}$ which describes the modulated Hamiltonian of the non-adiabatic process (Eqn. \ref{eq:dVecNon}) in the Pauli matrix space (with no on-site potential modulations).}
		\label{fig:Supp_kSpaceNOFM}
	\end{adjustwidth}
\end{figure}
This suggests the band structure is gapless, and confines the system to lie in the horizontal $\sigma_1-\sigma_2$ plane as shown in Fig. \ref{fig:Supp_kSpaceNOFM}. 
Calculating the Chern number for this process yields $\nu=0$. Fundamentally, this is the result of symmetry. Since in this non-adiabatic process the mirror symmetry is preserved, and has the following representation
\begin{eqnarray}\begin{aligned}
\hat{M}_x = \sigma_1,
\end{aligned}\end{eqnarray}
such that
\begin{eqnarray}\begin{aligned}
\hat{M}_x:H(k,\phi)\rightarrow H(-k,\phi)
\end{aligned}\end{eqnarray}
throughout the process. 
Consequently we have $d_i(k,\phi)\rightarrow d_i(-k,\phi)$ under the reflection symmetry. 
Calculating Eqn. \ref{eq:Chern_num_def}for this system, we find that the Chern number flips signs under reflection symmetry, and thus is always 0 if reflection symmetry is preserved. 
The transport of energy in this case may still occur due to Rabi like oscillations between the two coupled edge modes when the band gap shrinks or closes. This transport is therefore heavily dependent on timing as shown by the experimental results presented in Fig. 3 (main text), and Fig. \ref{fig:Supp_ccw}. 

%

%%%%%%%%%%%%%%%%%%%%%%%%%%%%%%%%%%%%%%%%%%%%%%%%%%%%%%%%%%%%%%%%%%%%%%%%%%%%%%%%%%%%%%%%%%%%%%%%%%%%%%%%%%%%%%%%%%%%%
%\newpage

\section{Discussion on adiabaticity}
\label{sec:Adiabaticity}

The adiabatic theorem, as it was originally proposed, states that a physical system remains in its instantaneous eigenstate if a given perturbation is acting on it slowly enough and if there is a gap between the eigenvalue and the rest of the Hamiltonian's spectrum \cite{Born1928}. 
In this section, we shall review the derivation of the adiabatic theorem and analytically establish the adiabatic limit for our dimerized array. We shall show that the calculated critical pump frequency, beyond which adiabaticity fails, agrees with our experimental result quantitatively.

We start by considering a general time-varying Hamiltonian $H(t)$, and its instantaneous eigenstates
\begin{eqnarray}\begin{aligned}
H(t)\psi_n(t) = \epsilon_n(t) \psi_n(t),
\label{eq:eigenstates}
\end{aligned}\end{eqnarray}
where the subscript $n$ indicates an eigenstate. 
For a generic process starting with $\psi_n(t=0)$ , after some time, typically the final state will not be $\psi_n(t)$ but rather a linear combination of all eigenstates. 
Therefore, for a process to be adiabatic, i.e. the $n^\text{th}$ eigenstate remains as the instantaneous state, it should have negligible probability to transition to any other state $\psi_m(t)$ in the spectrum. 
This means that the \emph{change} of $\psi_n(t)$ should have negligible overlap with any other state.
A mathematical description of this condition is therefore
\begin{eqnarray}\begin{aligned}
||\langle \psi_m(t)|\partial_t\psi_n(t)\rangle|| \ll 1,
\end{aligned}\end{eqnarray}
for all $m\neq n$ and at each time instance $t$, where $||\cdot||$ denotes the absolute value. 
Equivalently, by writing $|\psi_n(t+\delta t)\rangle = 
|\psi_n(t)\rangle + \delta t|\partial_t\psi_n(t)\rangle$, we get the following condition 
\begin{eqnarray}\begin{aligned}
\label{eq:inequality_2}
||\langle \psi_m(t)|\psi_n(t+\delta t)\rangle||  \ll \delta t,
\end{aligned}\end{eqnarray}
where we used the orthogonality of the eigenstates $\langle \psi_m(t)|\psi_n(t)\rangle=0$. 
Eqn.~\ref{eq:inequality_2} states that the overlap between $|\psi_m(t)\rangle$ and $|\psi_n(t+\delta t)\rangle$, which is at the next instance in time, should be much smaller than the time-step. 
In fact, such overlap is closely related to the energy gap between the two states. 
To see that, we substitute Eqn. \ref{eq:eigenstates} and rewrite Eqn. \ref{eq:inequality_2} as
\begin{eqnarray}\begin{aligned}
\left|\left|\frac{\langle\psi_m(t)|H(t)-H(t+\delta t)|\psi_n(t+\delta t)\rangle}{\epsilon_m(t)-\epsilon_n(t+\delta t)}\right|\right| \ll \delta t
\end{aligned}\end{eqnarray}
which we then rearrange as
\begin{eqnarray}\begin{aligned}
\quad \left|\left|\langle\psi_m(t)|\frac{H(t)-H(t+\delta t)}{\delta t}|\psi_n(t+\delta t)\rangle\right|\right| \ll  \left|\left|\epsilon_m(t)-\epsilon_n(t+\delta t)\right|\right|.
\end{aligned}\end{eqnarray}
By taking the limit $\delta t\rightarrow 0$ results in 
\begin{eqnarray}\begin{aligned}
\label{eq:inequality_3}
\left|\left|\langle \psi_m(t)|\dot{H}(t)|\psi_n(t)\rangle\right|\right|  \ll \left|\left|\epsilon_m(t)-\epsilon_n(t)\right|\right|,
\end{aligned}\end{eqnarray}
which recovers the standard adiabatic condition \cite{griffiths_schroeter_2018}. 
Intuitively, $\dot{H}(t)$ serves as the \emph{perturbation} that allows the transition between two instantaneously orthogonal states $\psi_{m,n}(t)$, which is otherwise forbidden for time-independent Hamiltonians. 
The value of this perturbation is closely related to the pump frequency $\omega_m$. This can be seen by taking the time derivative of the Hamiltonian given in Eqn. \ref{eq:HamiltonianReal} where the position in the pumping cycle is given by $\phi=\omega_m t$.
We can follow the state $\psi_n(t)$ adiabatically, given that the perturbation or the rate of change of the Hamiltonian, is smaller than the energy gap between $\psi_n(t)$ and \emph{any} other states. 

We note that Eqn.~\ref{eq:inequality_3} has to hold for all $m\neq n$ at all $t$. 
On the other hand, if there is a state that is degenerate with $\psi_n(t)$ at some time $t^*$, then the adiabatic theorem will break for $t= t^*$. 
This corresponds to the non-adiabatic pumping process discussed in the main text. 
In that experiment we turned off the frequency modulations and kept only the coupling modulations. During the pumping cycle when $\phi$ evolves from $0$ to $2\pi$ the band gap closes twice, as shown in Fig.~\ref{fig:Supp_Band} and Fig. \ref{fig:Supp_kSpaceNOFM}. As a result the whole process is non-adiabatic.

\begin{figure}[H]
	\begin{adjustwidth}{-0.5in}{-0.5in}
		\makebox[\textwidth][c]{\includegraphics[width=0.7\textwidth]{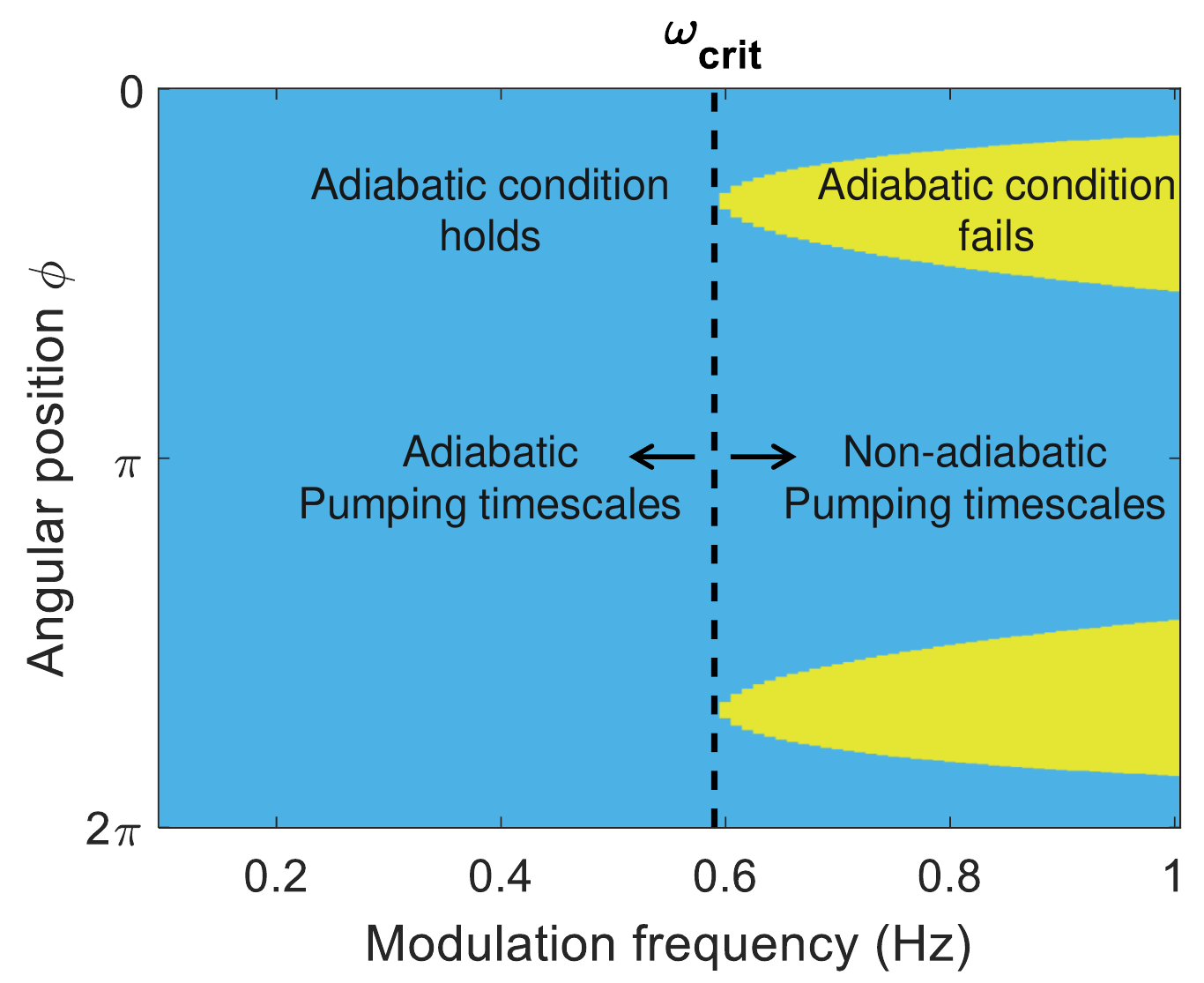}}
		\centering
		\caption{
			Simulated adiabatic condition given by $|| \langle\psi_m(t)|\dot{H}(t)|\psi_n(t)\rangle || - || \epsilon_m(t)-\epsilon_n(t) ||$ for parameters matching our experimental system. The blue region is where the difference is negative and the adiabatic condition holds, and the yellow region is where it does not. The dashed black line marks the critical pump frequency $\omega_\text{crit} \approx 0.6$ Hz, beyond which the the adiabatic condition fails for some phases $\phi$.}
		\label{fig:suppl_modfreq_adiabatic}
	\end{adjustwidth}
\end{figure}

We use parameters based on our topological pumping experiment and plot the condition in Eqn.~\ref{eq:inequality_3} for different values of $\omega_m$ in Fig. \ref{fig:suppl_modfreq_adiabatic}.
In the blue regions the value of perturbation is smaller than the energy gap and the condition in Eqn.~\ref{eq:inequality_3} holds, while in yellow regions it does not.
The lowest value of the pump frequency for which adiabaticity breaks ($\omega_\text{crit} \approx 0.6$ Hz), sets the critical limit for the process, separating between adiabatic and non-adiabatic regimes. 
This analytical result agrees very well with the experiments in the main text (Fig. 3) and in Fig.~\ref{fig:Supp_ccw}, where the normalized fractional energy $F_{L \rightarrow R}$ drops significantly around $\omega_m\approx0.6$ Hz.

%

%%%%%%%%%%%%%%%%%%%%%%%%%%%%%%%%%%%%%%%%%%%%%%%%%%%%%%%%%%%%%%%%%%%%%%%%%%%%%%%%%%%%%%%%%%%%%%%%%%%%%%%%%%%%%%%%%%%%%
\newpage

\section{Characterization of the experimental system}

\subsection{Experimental setup}
\label{sec:suppl_exp_setup}

\begin{figure}[h]
	\begin{adjustwidth}{-0.5in}{-0.5in}
		\makebox[\textwidth][c]{\includegraphics[width=1\textwidth]{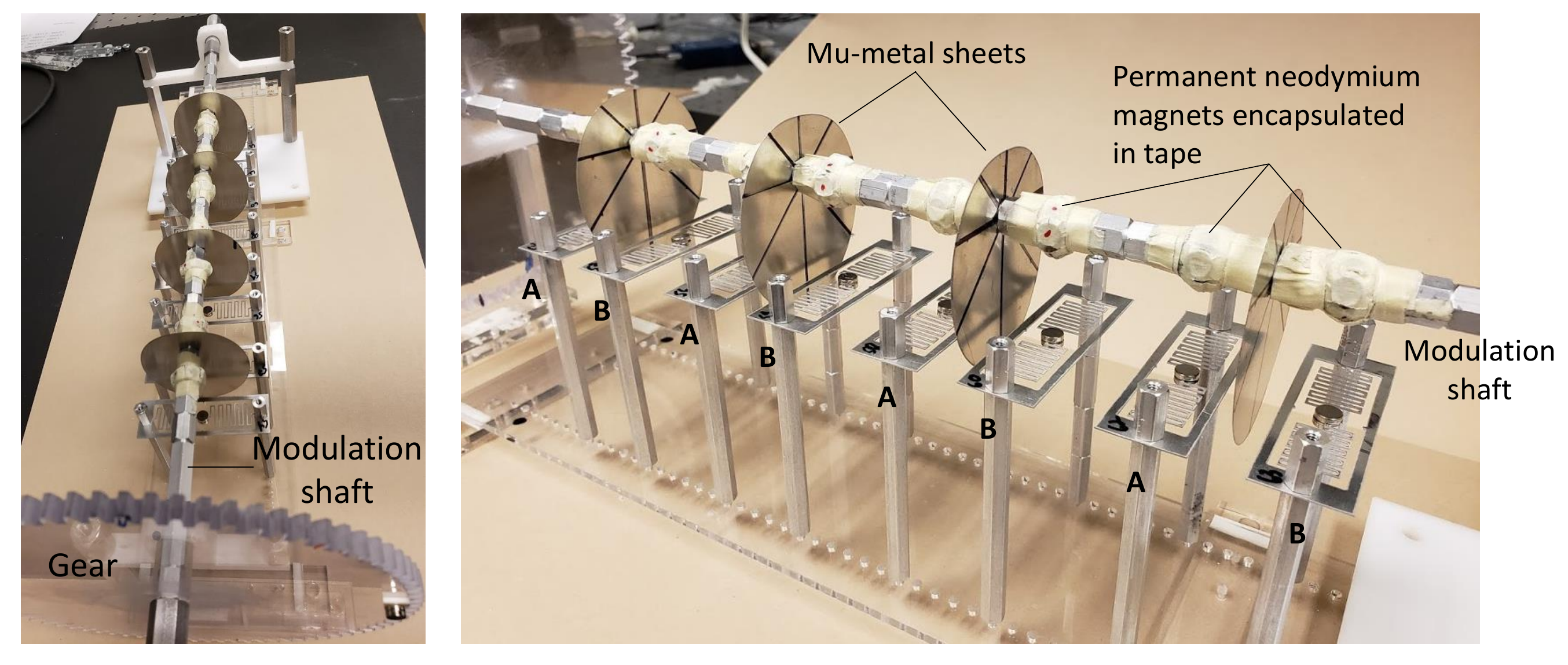}}
		\centering
		\caption{
			Photograph of the 8 resonator array experimental setup. 	
		}
		\label{fig:Supp_Setup}
	\end{adjustwidth}
\end{figure}

The experimental setup composed of an 8 resonator array is shown in Fig. \ref{fig:Supp_Setup}. The modulation shaft is positioned above the resonators and is connected to a motor at one end and a bearing at the other end to allow free rotation. Four mumetal sheets are connected to the shaft and positioned between A and B sites in each unit-cell  to function as the coupling modulators.
A series of permanent neodymium (N52 material) are glued to the shaft at the resonator positions and induce the required on-site frequency modulations. 
Each resonator is equipped with a Hall sensor (not shown) to measure the change in magnetic field and infer angular displacement.

\subsection{Resonator characterization}
\label{sec:suppl_res_char}

The typical -3 dB bandwidth of our mechanical resonators is $\Delta f \approx 0.38$ Hz, which results in a decay time of the mechanical vibrational energy of $\tau = 1/\pi \Delta f \approx 0.85$ sec. 
The adiabatic limit calculated for our system in \S~\ref{sec:Adiabaticity} is $0.6$ Hz, implying that the pumping cycle time must be $1.7$ sec or longer. 
This means that the resonator decay timescale is not sufficient for convenient experimental observation of the pumping process.

\begin{figure}[H]
	\begin{adjustwidth}{-0.5in}{-0.5in}
		\makebox[\textwidth][c]{\includegraphics[width=1\textwidth]{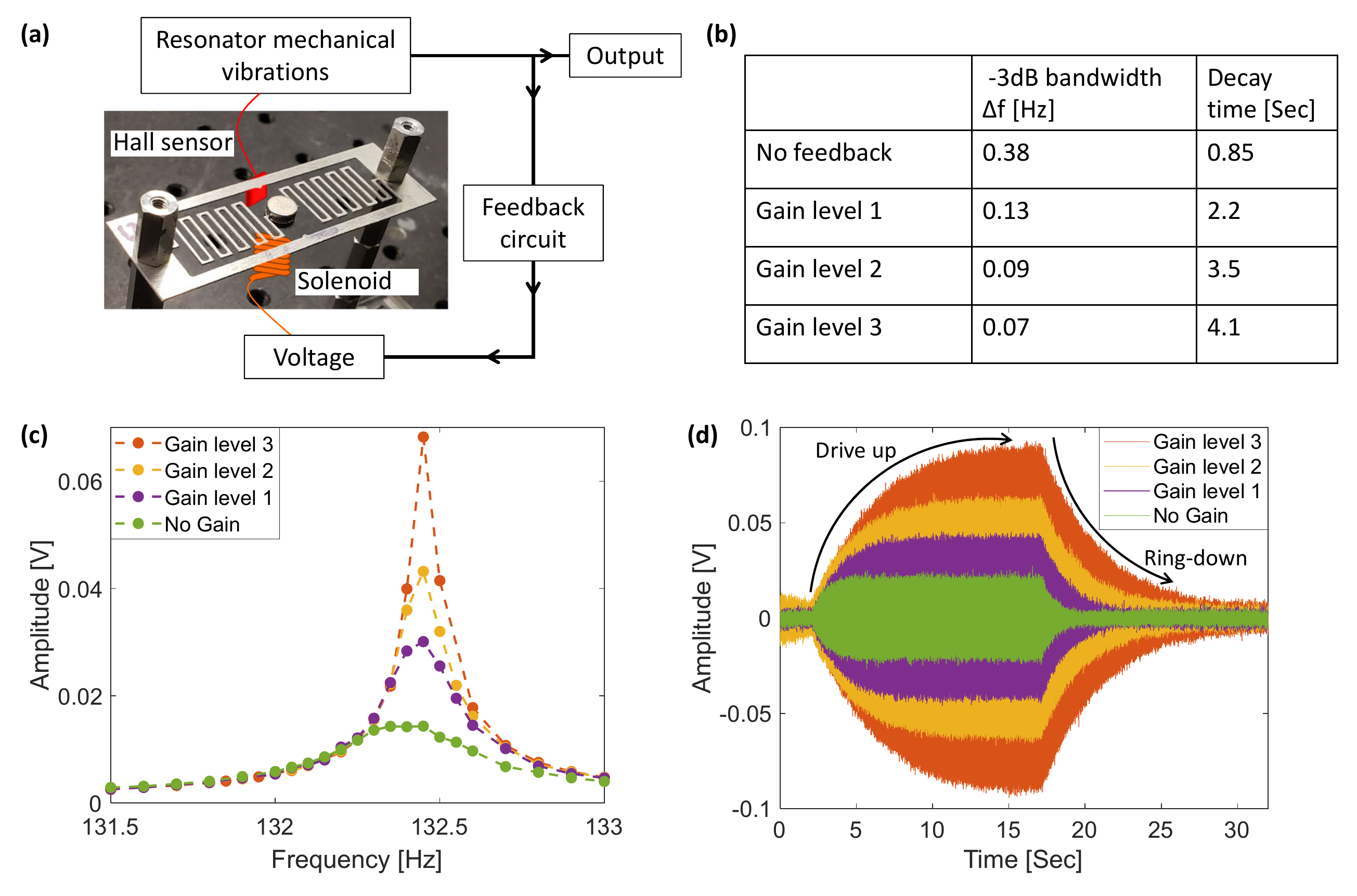}}
		\centering
		\caption{
			(a) Schematic of the anti-damping feedback system. (b) Table summarizing the -3dB bandwidth and decay times for a resonator without feedback and with three different feedback settings. (c)	The frequency response of a resonator without feedback and with the three different feedback settings and (d) ring down measurements when excited at the resonance frequency.
		}
		\label{fig:Supp_Gain}
	\end{adjustwidth}
\end{figure}

In order to increase the decay time of the resonators, we implemented a feedback anti-damping circuit for each resonator (Fig. \ref{fig:Supp_Gain}a). 
The output voltage from the Hall sensor of each resonator (equivalent to angular displacement $\theta$) is fed back after amplification and a $\pi/2$ phase shift (equivalent to angular velocity $\dot{\theta}$) to a compact solenoid coil adjacent to the resonator. Since this force feedback is proportional to $\dot{\theta}$, it can reduce the action of the viscous damping $c$ (in Eqn. \ref{eq: Second_Order}) and increase the effective Q-factor and decay time of the resonator. The remainder of the resonator dynamics e.g. frequency, remain unchanged.

Typical results of a resonator without feedback are compared to three different feedback settings in Fig. \ref{fig:Supp_Gain}c. The corresponding values of -3dB bandwidth and decay times are reported in the table in Fig. \ref{fig:Supp_Gain}b.
In the experiments presented in the main text we choose to apply the feedback setting of level 2 to the resonators to set a decay time of ~$3.5$ sec.

\newpage
\subsection{Frequency modulation}
\label{sec:suppl_frequency_modulation}

\begin{figure}[hp]
	\begin{adjustwidth}{-0.5in}{-0.5in}
		\makebox[\textwidth][c]{\includegraphics[width=0.7\textwidth]{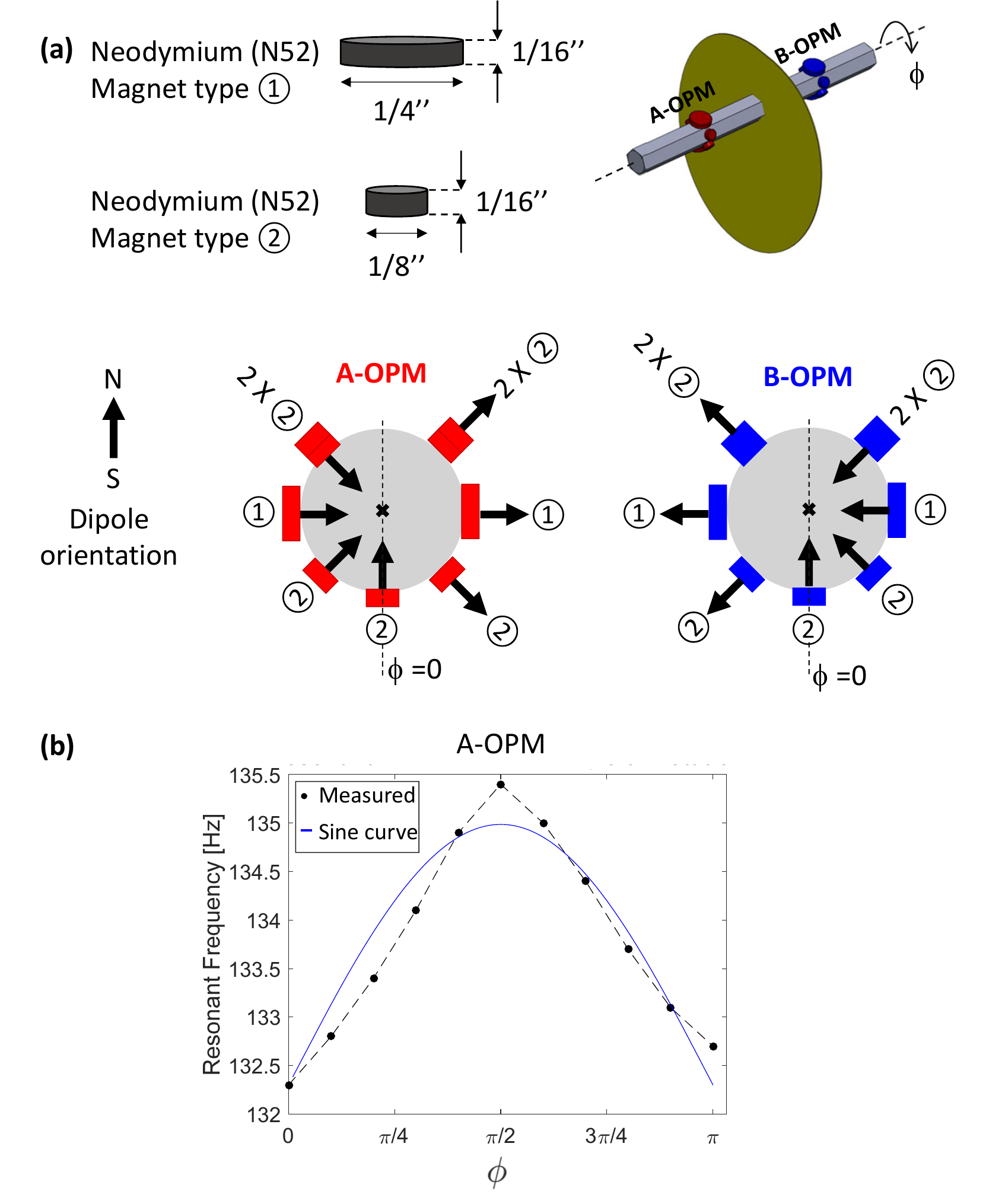}}
		\centering
		\caption{
			(a) Illustrations of the magnet arrangement on the modulation shaft circumference used for on-site potential (frequency) modulations. Two different sizes of magnet were used and their arrangement for site A (A-OPM) and B (B-OPM) is with opposite phase. (b) Measured values of the resonance frequency of site A as a function of the modulation angle $\phi$.  	
		}
		\label{fig:Supp_OPM}
	\end{adjustwidth}
\end{figure}

We achieve frequency modulations in the resonator array, through the magnetostatic spring effect (see \S~S1, Eqn \ref{eq: Second_Order}). 
A sequence of permanent magnets are attached to the circumference of the shaft above each resonator to function as the on-site potential modulator (OPM).
As the shaft rotates, different magnets come into proximity with the resonator and induce an on-site frequency shift.
The $A$ and $B$ sites are modulated with magnets of the same magnitude but opposite phasing as shown in Fig. \ref{fig:Supp_OPM}a. 
Measured values of resonance frequency of site A as a function of the angular position $\phi$ are presented in Fig. \ref{fig:Supp_OPM}b.

\newpage
\subsection{Coupling modulations}
\label{sec:suppl_coupling_modulation}

\begin{figure}[hp]
	\begin{adjustwidth}{-0.5in}{-0.5in}
		\makebox[\textwidth][c]{\includegraphics[width=0.7\textwidth]{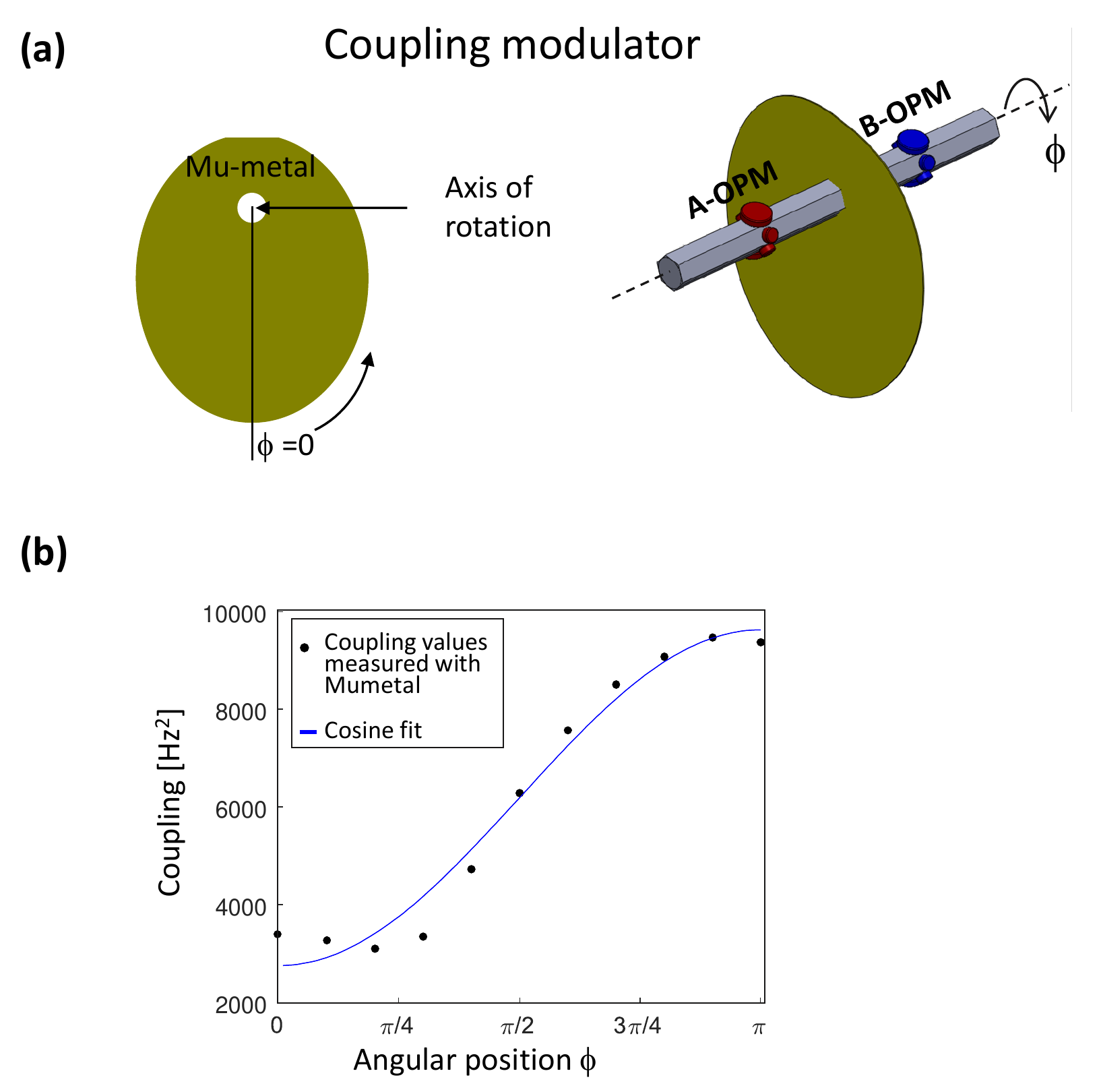}}
		\centering
		\caption{
			(a) Design of the mu-metal coupling modulator to achieve the desired cosine modulation. (b) Experimentally measured coupling rate between resonators as function of the angular position of the modulation shaft.
		}
		\label{fig:Supp_Mumetal}
	\end{adjustwidth}
\end{figure}

In order to modulate the coupling between resonators in a unit cell we use mu-metal (ferromagnetic material with high permeability) sheets, that divert the magnetic field between the resonators and reduce the magnetic coupling.
The shape of this coupling modulator (Fig. \ref{fig:Supp_Mumetal}a) was determined through experimental iterations. To measure the coupling rates experimentally we use a two resonator setup and evaluate the mode splitting.
The measured coupling rates as a function of the angular position of the mu-metal sheet $\phi$ are shown in Fig. \ref{fig:Supp_Mumetal}b.

%%%%%%%%%%%%%%%%%%%%%%%%%%%%%%%%%%%%%%%%%%%%%%%%%%%%%%%%%%%%%%%%%%%%%%%%%%%%%%%%%%%%%%%%%%%%%%%%%%%%%%%%%%%%%%%%%%%%%
\newpage

\subsection{Band structure}
\label{sec:suppl_Band_structure}

In this section we will present simulated and measured band-structures of our system, for both the topological pumping and for the non-adiabatic process described in the main text and \S S2. 
We will discuss some effects that influence our experimental system that are not included in the ideal model used to calculate the band structure presented in Fig. 1b in the main text.

The topological pump is described by the Hamiltonian in Eqn. \ref{eq:HamiltonianReal}, where resonance frequencies (on-site potentials) as well as the coupling rates are being modulated. 
In this case the band-gap does not close during the pumping cycle. As long as the adiabatic condition (\S \ref{sec:Adiabaticity}) is satisfied, it is guaranteed that we follow the same eigenstate from one edge of the array to the other, and transport the vibrational energy.
A non-adiabatic process is demonstrated by a system where only the coupling rates are modulated. In this process the band gap closes twice in a cycle and therefore the adiabatic window collapses. In such a process, energy may oscillate between the two degenerate edge modes in a manner similar to Rabi oscillations. 
The two different mechanisms are explained in the main text and illustrated in Fig. 3 as well as in Fig. \ref{fig:Supp_kSpaceFM} and Fig. \ref{fig:Supp_kSpaceNOFM} in \S\ref{sec:suppl_Cal_Chern}.

\begin{figure}[H]
	\begin{adjustwidth}{-0.5in}{-0.5in}
		\makebox[\textwidth][c]{\includegraphics[width=1.1\textwidth]{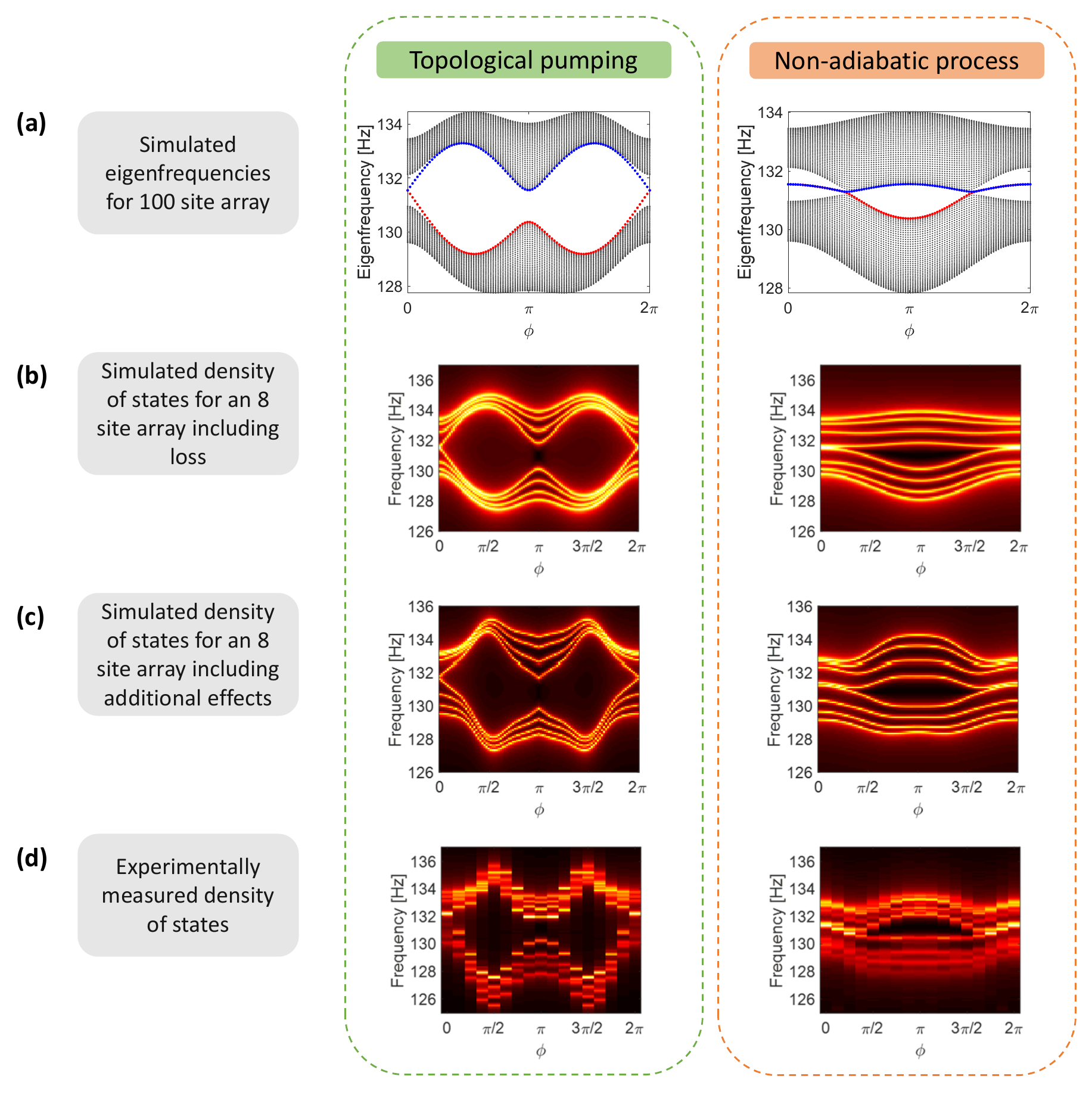}}
		\centering
		\caption{
			Simulated and measured eigenfrequencies of the 1D modulated array for both the topological pump (left) and the non-adiabatic process (right). 
			(a) Simulated eigenfrequencies as a function of the pumping parameter $\phi$ for an ideal array of 100 sites. The lower (EM$_\text{L}$) and upper (EM$_\text{U}$) paths of the edge modes are colored red and blue respectively. 
			(b) Simulated eigenfrequencies of an 8 site array including intrinsic resonator loss. Brighter color represents greater density of states. 
			(c) Simulated eigenfrequencies of an 8 site array including loss, next-nearest neighbor coupling, and frequency and coupling modulations based on experimental calibration. 
			(d) Experimentally measured band-structure as a function of the pump angular position $\phi$. Due to the symmetry of the system the measurements were taken in the range $\phi=[0,\pi]$ and mirrored for the plot.  	
		}
		\label{fig:Supp_Band}
	\end{adjustwidth}
\end{figure}

In Fig. \ref{fig:Supp_Band} we present plots of the eigenfrequencies of the 1D array throughout the pumping process.
We first simulate the eigenvalues for a 100 site (50 unit cells) lossless system as a function of the pumping parameter $\phi$ (Fig. \ref{fig:Supp_Band}a).
At $\phi=0,2\pi$ the system is in the topologically non-trivial phase with two degenerate edge modes within the bulk band gap. 
In the topological pump the degeneracy of these edge modes is lifted for $\phi>0$ due to the frequency modulations which break inversion symmetry, and the bandgap remains open throughout the pump cycle. 
In the non-adiabatic process the two edge modes stay degenerate until the bandgap closes, which happens twice during the pumping cycle.

Next, we wish to simulate approximately the band structure of our experimental system.
We begin by simulating an array of 8 sites, and include a loss parameter evaluated based on experimental measurements (Fig. \ref{fig:Supp_Band}b). 
For a system which includes loss we can no longer calculate real eigenvalues. We therefore simulate the mechanical density of states (equivalent to mechanical susceptibility defined as the torque-to-angular-displacement transfer function) of the array at each value of $\phi$. Repeating this process for values of $\phi$ in the range of $[0,2\pi]$ visualizes the band structure of the system. 

In this simulation we observe a similar trend as the case with 100 sites. 
We now include a few additional effects that are inevitable in any experimental system, in the simulations shown in Fig. \ref{fig:Supp_Band}c. The first is next nearest neighbor coupling which we estimate based on measured values and the cubic decay of the magnetic coupling with distance. A second effect is that the modulations do not follow a perfect sinusoidal curves. We experimentally extract modulation functions based on fits to the measured values (see \S \ref{sec:suppl_frequency_modulation} and \S \ref{sec:suppl_coupling_modulation}). 
These effects change the band structure slightly and better approximate the actual experimental measurements which are shown in Fig. \ref{fig:Supp_Band}d.

\newpage

\begin{figure}[h]
	\begin{adjustwidth}{-0.5in}{-0.5in}
		\makebox[\textwidth][c]{\includegraphics[width=0.9\textwidth]{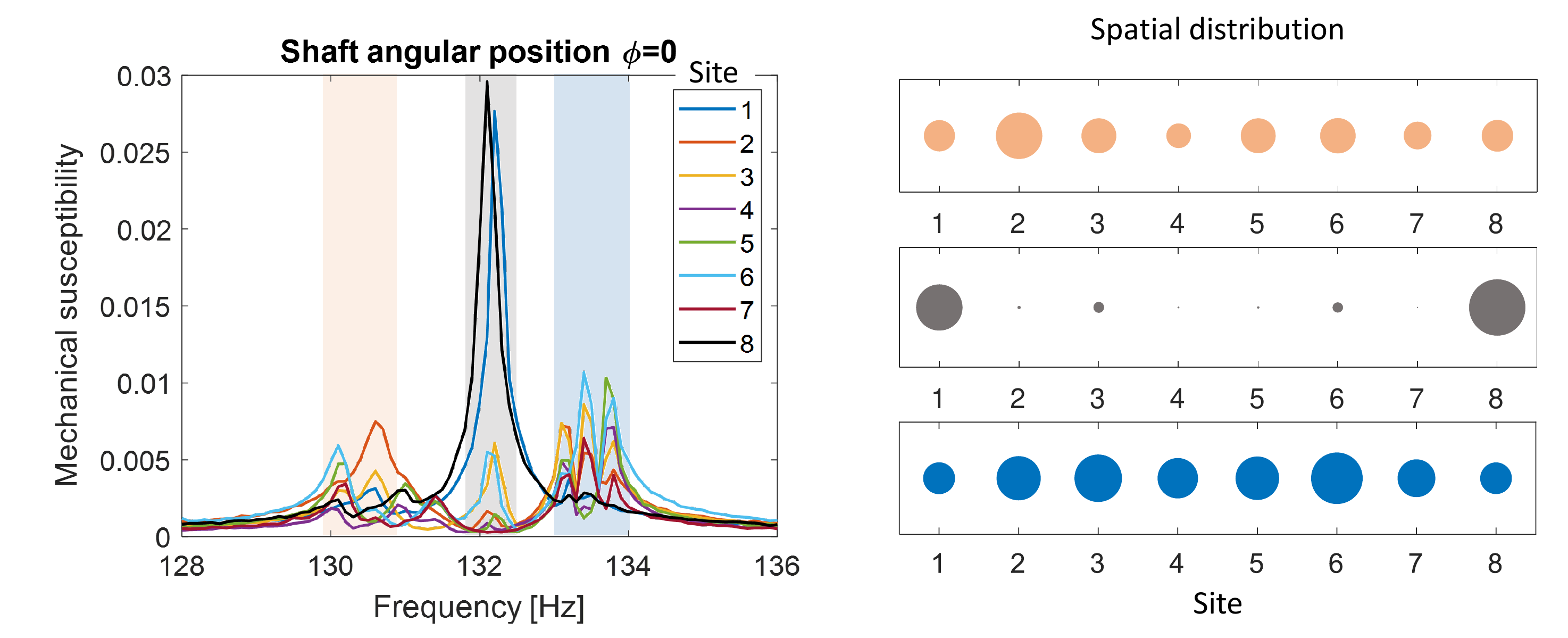}}
		\centering
		\caption{
			Characterization of the array at $\phi=0$. The experimentally measured spectrum of all 8 resonators is presented on the left, showing the mechanical susceptibility (ratio of angular displacement amplitude to applied torque, in arbitrary units) as a function of the drive frequency at each resonator. These measurements were performed by locally exciting each resonator with a small solenoid coil and measuring its mechanical response. The spatial distribution of the energy over the three frequency ranges highlighted on the spectrum plot is shown on the right. Circle size corresponds to energy magnitude averaged over the highlighted frequency range for each site.
		}
		\label{fig:Supp_spec_phi0}
	\end{adjustwidth}
\end{figure}

The experimentally measured spectrum of all 8 resonator sites at $\phi=0$ is presented in Fig. \ref{fig:Supp_spec_phi0}. Resonators 1 and 8 on the edges of the array show a prominent mode localized within the bulk band gap. The spatial distribution of the integrated energy (angular oscillation amplitude squared) is shown on the right panel of Fig. \ref{fig:Supp_spec_phi0} where circle size corresponds to magnitude. While for the lower and upper bands the distribution is almost uniform throughout the array, for the mid gap frequency range the energy is strongly localized at the two edges.

%\newpage
\subsection{Eigenmode decomposition}
\label{sec:suppl_eigenmode_decomp}

A typical measurement from our pumping experiment as presented in Fig. 2e of the main text includes the vibrational motion of each resonator in the array as a function of time.
The harmonic displacement $x_n(t)$ of the $n^\text{th}$ resonator can be written as the superposition of the system's eigenmodes such that $x_n(t)=\sum_{m=1}^{N}\psi_\text{mn}a_m(t)$. Here $\psi_\text{mn}$ is the shape contribution of the $m^{th}$ eigenmode at the $n^\text{th}$ resonator (i.e. components of the eigenvectors), and $a_m$ is its amplitude.
By taking the Fourier transform of this equation we find the frequency domain expression $X_n(\omega)=\sum_{m=1}^{N}\psi_\text{mn}A_m(\omega)$. 
Here $X_n(\omega)$, $A_m(\omega)$ corresponding spectra including both amplitude and phase information.
We can write this relationship in the matrix form $X=\Psi A$ where $X$ is a column vector $[X_1(w), X_2(w) ... ]^T$ and $A$ is a column vector $[A_1(w), A_2(w) ...]^T$. We can then extract the eigenmodes spectra from the measured displacement spectra using the inverse relation $A=\Psi^\text{-1}X$. 
Finally, the square of the spectrum $A^2$ is equivalent to the vibrational energy in each eigenmode. 
We define an energy fraction for each eigenmode ($E_\text{mode \#}$) as the fraction of total mechanical energy in the array projected onto the selected mode.
We repeat this computational process for overlapping time segments of $0.25$ sec throughout the pumping cycle, and track the energy in the different eigenmodes throughout the process. 
A typical result of this analysis is shown in Fig. 2g of the main text. 

\subsection{Simulations of the transport fidelity}
\label{sec:suppl_Simulaitons}

\begin{figure}[H]
	\begin{adjustwidth}{-0.5in}{-0.5in}
		\makebox[\textwidth][c]{\includegraphics[width=0.9\textwidth]{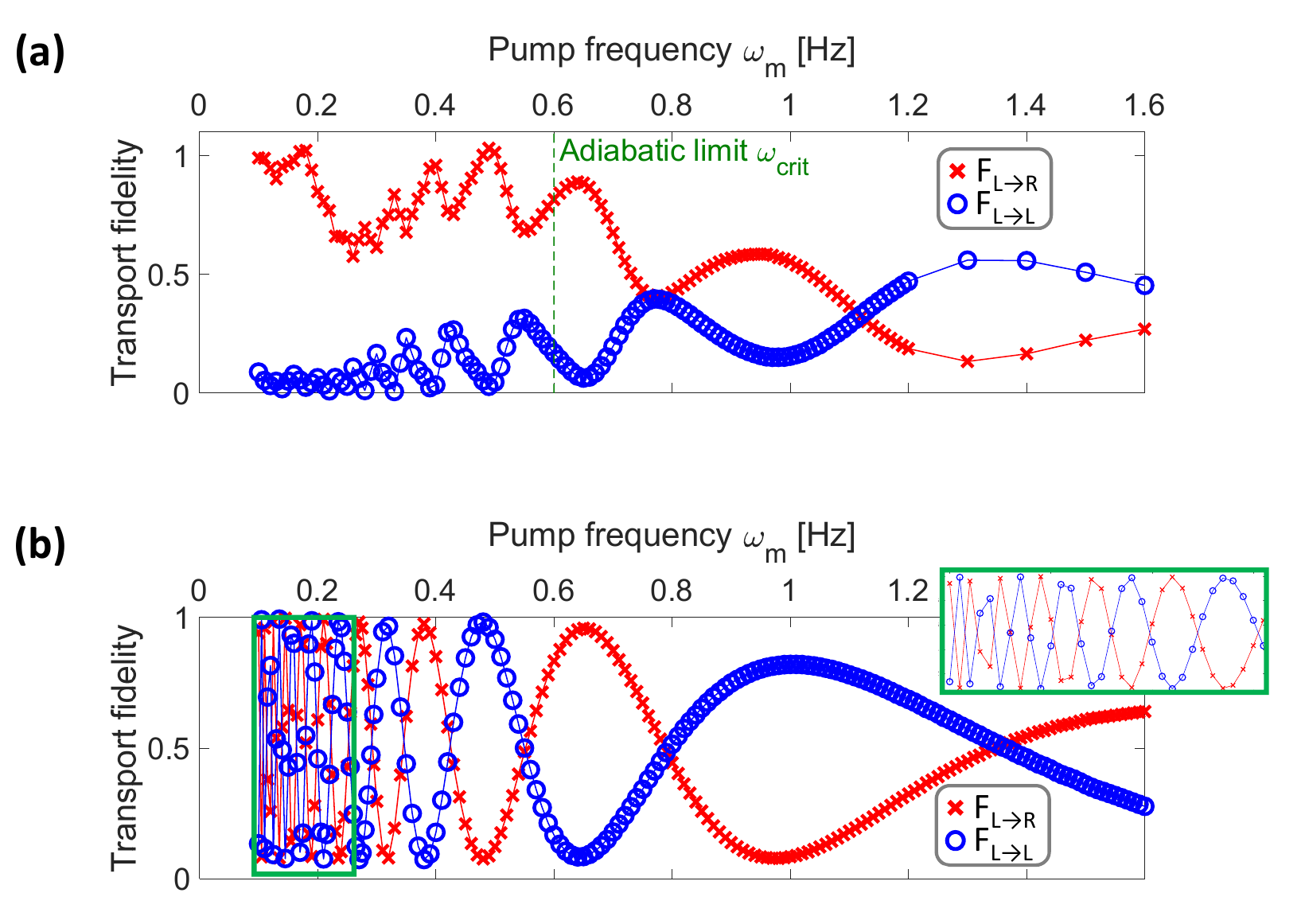}}
		\centering
		\caption{
			Transport fidelity derived from simulations of an 8 resonator system as a function of the pump frequency $\omega_m$. (a) Simulation of a topological pump and (b) of a non-adiabatic process. 
		}
		\label{fig:Supp_FidSim}
	\end{adjustwidth}
\end{figure}

In this subsection we present simulation results of the transport fidelity values $F_{L \rightarrow R}, F_{L \rightarrow L}$.
We simulate both the topological pump in which both frequencies and coupling values are modulated as well as the limit case of a non-adiabatic process where only coupling values are modulated, as discussed in the main text. 
To produce these simulations we first use a time domain solver for the full nonlinear equations of motion of the 8 resonator array. The resulting vibrational motion of all resonators is obtained, similar to the data we obtain experimentally (Fig. 2e). 
We then repeat the eigenmode decomposition process explained in \S~\ref{sec:suppl_eigenmode_decomp}, and calculate the transport fidelity values as defined in the main text.
This simulation is repeated for many different pump frequencies, and the results are presented in Fig. \ref{fig:Supp_FidSim}. 
For topological pumping the energy is reliably transported from the left edge to the right edge of the array for a range of pump frequencies up to a critical value $\omega_\text{crit}$. 
In contrast, for the non-adiabatic process, energy oscillates between the two edges due to Rabi like oscillations. At the end of the non-adiabatic cycle the energy can be localized at either edge and is heavily dependent on timing, as shown by the oscillating values of $F_{L \rightarrow R},F_{L \rightarrow L}$.  

%%%%%%%%%%%%%%%%%%%%%%%%%%%%%%%%%%%%%%%%%%%%%%%%%%%%%%%%%%%%%%%%%%%%%%%%%%%%%%%%%%%%%%%%%%%%%%%%%%%%%%%%%%%%%%%%%%%%%
%\newpage

\section{Additional experimental results}

\subsection{ccw pumping experiments}
\label{sec:suppl_exp_ccw}

\begin{figure}[H]
	\begin{adjustwidth}{-0.5in}{-0.5in}
		\makebox[\textwidth][c]{\includegraphics[width=0.8\textwidth]{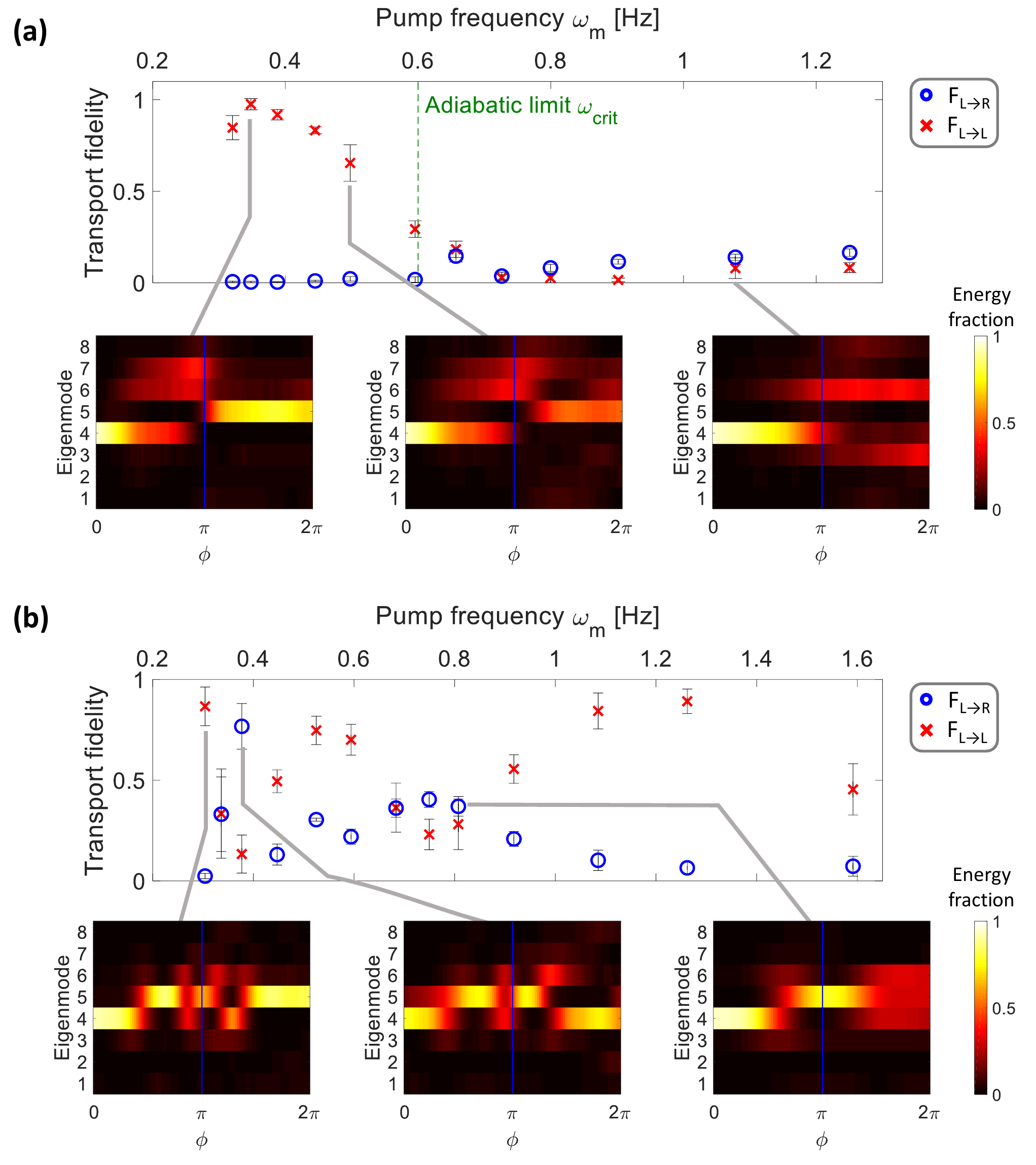}}
		\centering
		\caption{
			Experimentally measured transport fidelity values for ccw rotation of the modulation shaft, for both (a) the topological pump (pumping along EM$_\text{U}$) and (b) the non-adiabatic process. Each data point is averaged over 10 consecutive measurements and the error bars represent the standard deviation amongst these measurements. The insets are examples of the modal energy fraction throughout the pumping cycle.
		}
		\label{fig:Supp_ccw}
	\end{adjustwidth}
\end{figure}

In the main text we presented experimental results of the measured transport fidelity values $F_{L \rightarrow R}, F_{L \rightarrow L}$, for the topological pump and the non-adiabatic process for clockwise (cw) rotation of the modulation shaft. 
Here we present experimental results for ccw rotation of the modulation shaft (Fig. \ref{fig:Supp_ccw}) which shows similar trends. 
The difference between cw and ccw rotations can be understood from the band-structure shown in Fig. \ref{fig:Supp_Band}a. For cw topological pumping we follow the lower edge mode $\text{EM}_\text{L}$, while when it rotates ccw we follow the upper edge mode $\text{EM}_\text{U}$. Both yield the same outcome with the edge modes being transported from one side of the array to the other.

%%%%%%%%%%%%%%%%%%%%%%%%%%%%%%%%%%%%%%%%%%%%%%%%%%%%%%%%%%%%%%%%%%%%%%%%%%%%%%%%%%%%%%%%%%%%%%%%%%%%%%%%%%%%%%%%%%%%%
%\newpage

\subsection{Experimental results for different defects}
\label{sec:suppl_exp_defects}
In this section we present additional experimental results that for the sake of brevity were not included in the main text. 

\begin{figure}[H]
	\begin{adjustwidth}{-0.5in}{-0.5in}
		\makebox[\textwidth][c]{\includegraphics[width=0.7\textwidth]{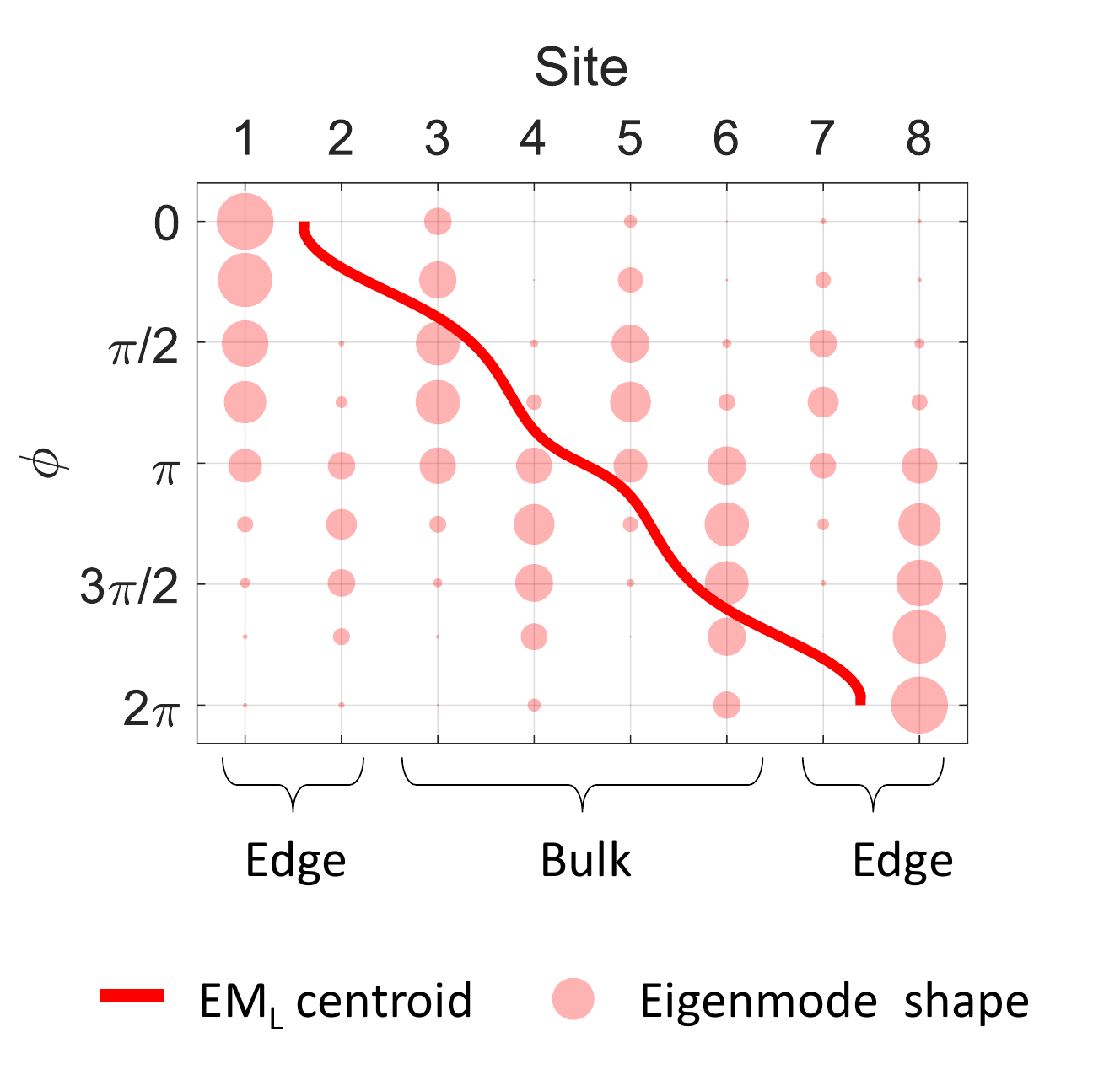}}
		\centering
		\caption{
			Eigenvector spatial distribution for different values of $\phi$ in increments of $\pi/4$ are shown in shaded pink, where circle size corresponds to magnitude of the local excitation. The solid red line is the centroid of the eigenvector and helps visualize the spatial evolution of the mode across the array. This plot is for a system without any defects and is used as a visual aid to understand the spatio-temporal coordinates of the temporal defects in Figs \ref{fig:Supp_Defect_Temporal1} and \ref{fig:Supp_Defect_Temporal2} and in Fig. 5 in the main text.}
		\label{fig:Supp_Centroid}
	\end{adjustwidth}
\end{figure}

In Fig. \ref{fig:Supp_Centroid} we show the spatial distribution of $\text{EM}_\text{L}$ for $\pi/4$ spacing of $\phi$ and its centroid throughout the pumping process for an unperturbed system. 

\begin{figure}[H]
	\begin{adjustwidth}{-0.5in}{-0.5in}
		\makebox[\textwidth][c]{\includegraphics[width=1.2\textwidth]{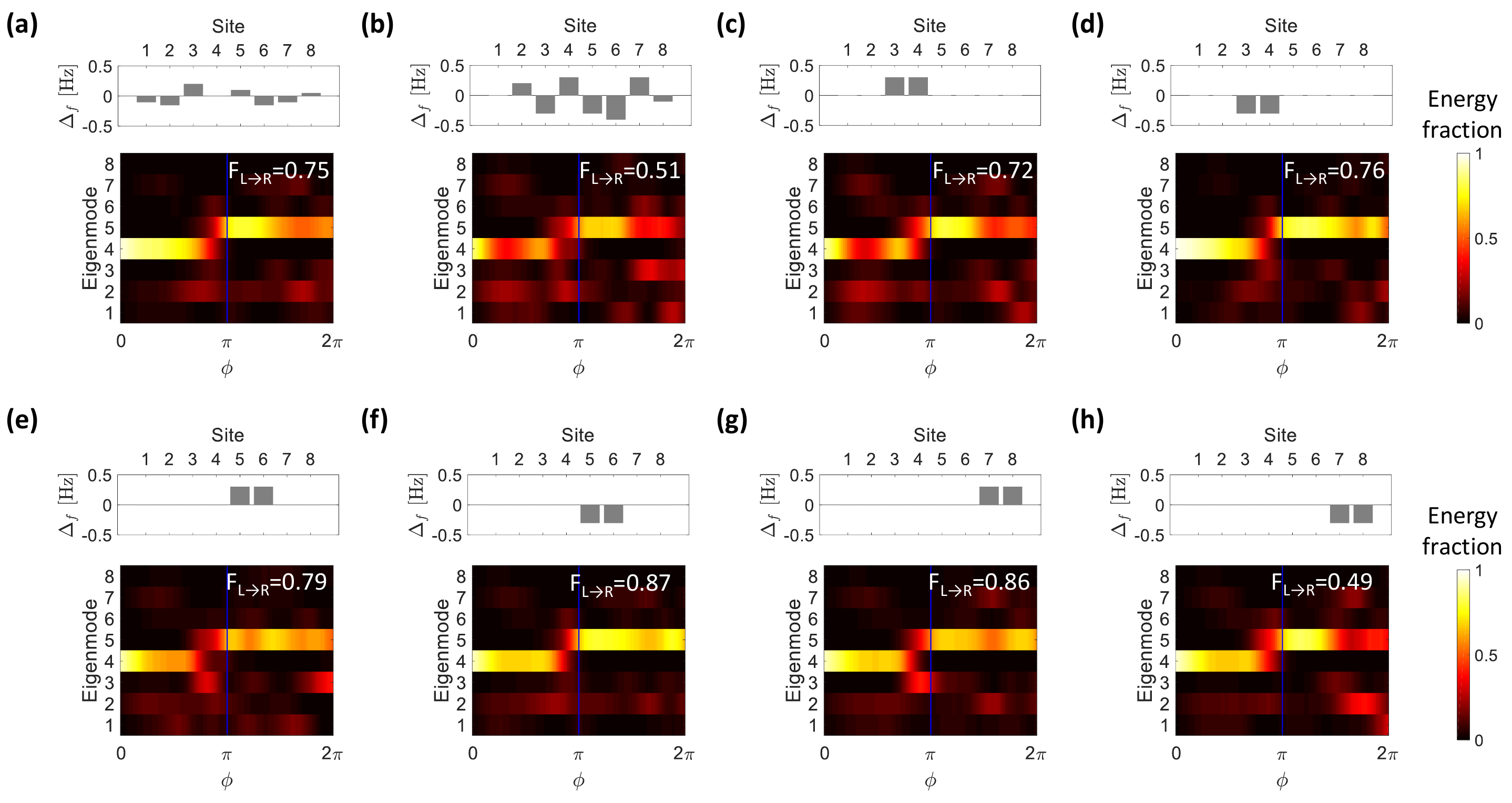}}
		\centering
		\caption{
			Modal energy fraction for a system with spatial on site potential defects.
			The bar plots above each modal energy fraction indicates the amount of detuning applied at each site. All of the results are for modulation rate of $\omega_m \approx 0.3$ Hz.
			In (a) the resonance frequencies were randomly shifted in the range of $\pm0.2$  Hz which is $\sim$10\% of the band-gap.
			In (b) the resonance frequency of the array was randomly shifted in the range of $\pm0.4$ Hz which is $\sim$20\% of the system band-gap. In this case the array disorder is very large and the pump efficiency dropped to 50\%. 
			In (c-h) the resonance frequency of a single unit cell was shifted by either $\pm 0.3$ Hz. When the defect is at the bulk unit cells (i.e. resonators 3-6) the transport fidelity remains high showing the system robustness.
		}
		\label{fig:Supp_Defect_Stat}
	\end{adjustwidth}
\end{figure}

Figure \ref{fig:Supp_Defect_Stat} presents additional experimental results of the type shown in Fig. 4 of the main text. In these experiments we incorporate static defects of on-site frequency detuning. We also include two examples where pumping ocurrs with lo fidelity $F_{L\rightarrow R}$. 

Figure \ref{fig:Supp_Defect_Temporal1} and \ref{fig:Supp_Defect_Temporal2} present additional experimental results on temporal defects of the type shown in Fig. 5a of the main text. Here we incorporate a temporal on-site frequency detuning for different sites in the array and at different angular positions of the modulation shaft.

Finally, in Fig. \ref{fig:Supp_Defect_PhaseBoundary} we present a visual explanation of the phase boundary defect presented in Fig. 5b of the main text.

\begin{figure}[H]
	\begin{adjustwidth}{-0.5in}{-0.5in}
		\makebox[\textwidth][c]{\includegraphics[width=1.1\textwidth]{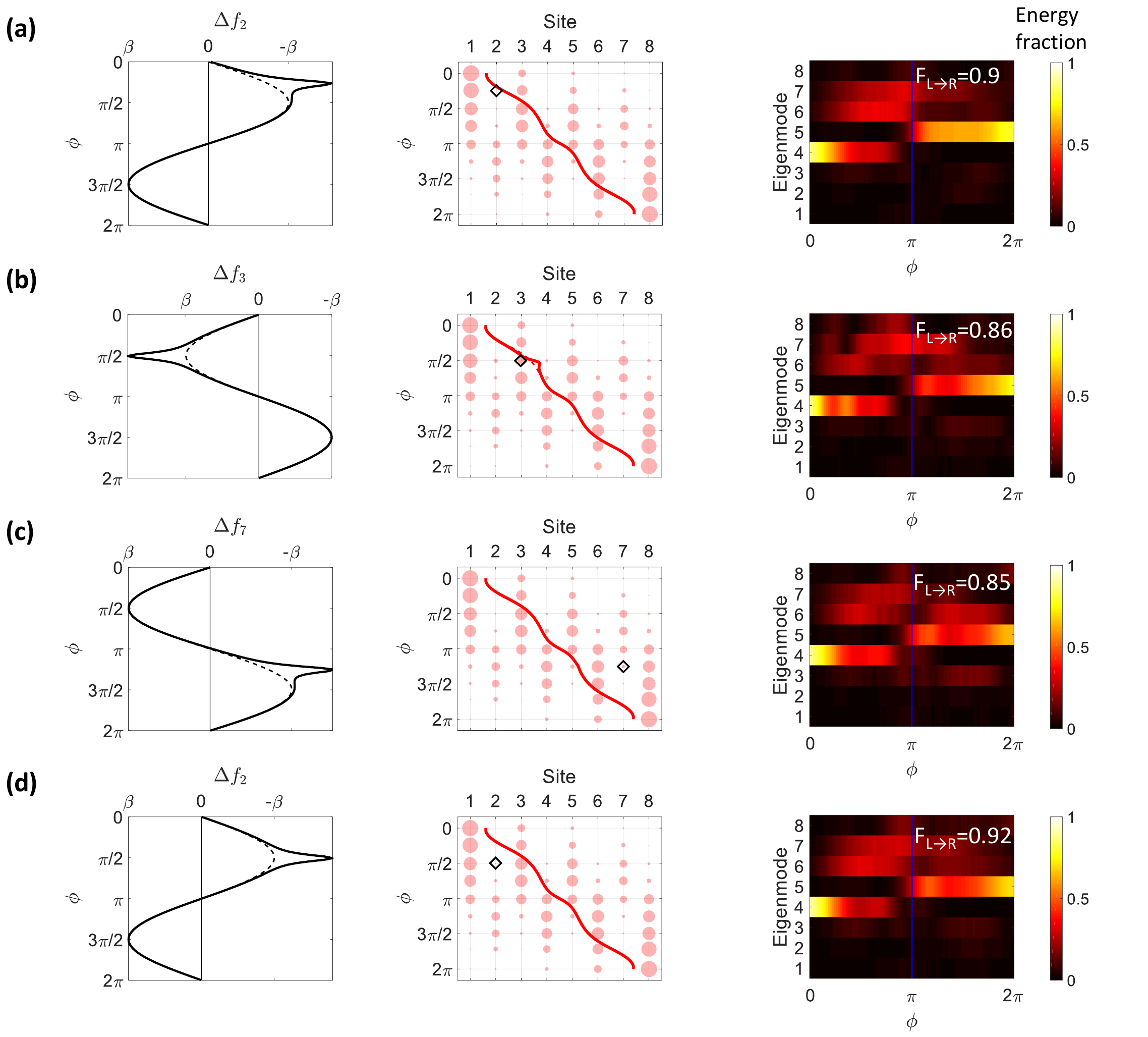}}
		\centering
		\caption{
			Modal energy fraction throughout the pumping process for a system with different spatio-temporal defects of on-site potential, and modulation rate of $\omega_m \approx 0.3$ Hz. 
			For each result we show schematically the evolution of the resonance frequency of the specific defect resonator $\Delta f_{\#}$. The simulated evolution of the \EML $\,$ eigenvector and centroid are also presented (see comparison to an unperturbed system in Fig. \ref{fig:Supp_Centroid}). The coordinates of the spatio-temporal defect are marked by a diamond on the centroid plot. Examples continue in Fig \ref{fig:Supp_Defect_Temporal2}.
		}
		\label{fig:Supp_Defect_Temporal1}
	\end{adjustwidth}
\end{figure}

\begin{figure}[H]
	\begin{adjustwidth}{-0.5in}{-0.5in}
		\makebox[\textwidth][c]{\includegraphics[width=1.1\textwidth]{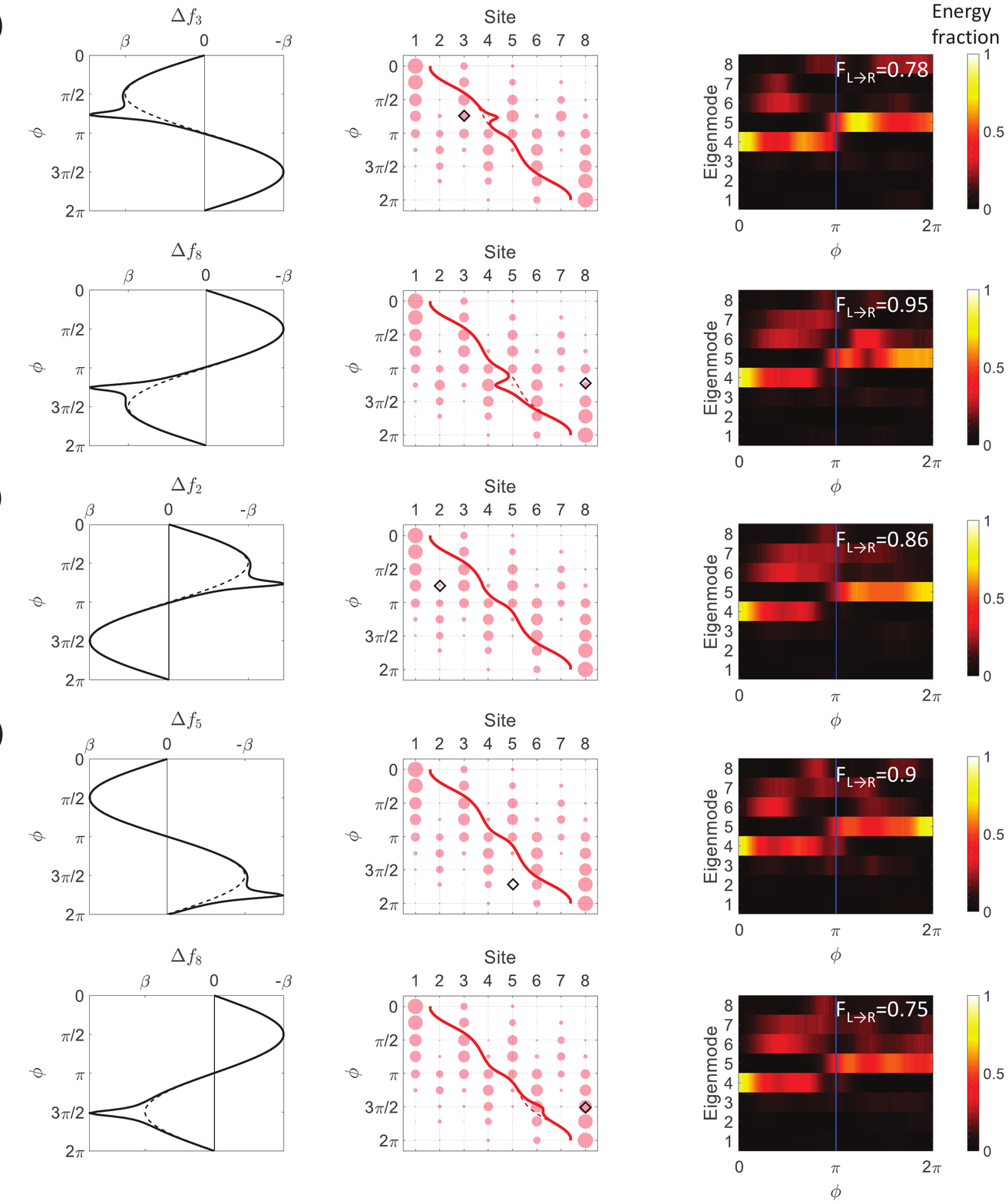}}
		\centering
		\caption{
			Continuation of spatio-temporal defects from Fig. \ref{fig:Supp_Defect_Temporal1}.
		}
		\label{fig:Supp_Defect_Temporal2}
	\end{adjustwidth}
\end{figure}

\begin{figure}[H]
	\begin{adjustwidth}{-0.5in}{-0.5in}
		\makebox[\textwidth][c]{\includegraphics[width=1.2\textwidth]{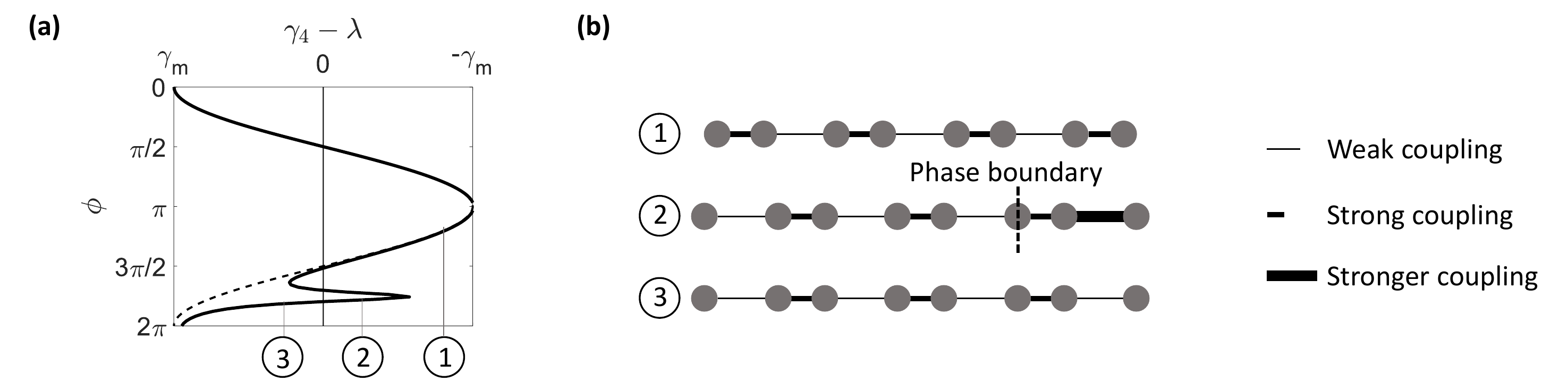}}
		\centering
		\caption{
			A visual explanation of the phase boundary defect presented in Fig. 5b of the main text. 
			(a) The intracell coupling rate at the 4$^{\text{th}}$ unit cell was altered from the standard trajectory (as shown by the $\gamma_4-\lambda$ plot) near the end of the pump cycle. Three points of interest are identified for closer examination, indicated by (1), (2), and (3). 
			(b) At these points, we visualize the tight-binding form (ignoring on-site potentials) with line width corresponding to the coupling strength. At point (2) the system exhibits the intrusion of a phase boundary between trivial and topological phases. 
		}
		\label{fig:Supp_Defect_PhaseBoundary}
	\end{adjustwidth}
\end{figure}

\bibliographystyle{naturemag}
%\bibliography{mmti_pump}

\end{document}